\documentclass[oldversion]{aa}
\usepackage{amsmath}
\usepackage{graphicx}
\usepackage{color}
\usepackage{txfonts}
\usepackage{hyperref}
\usepackage{cleveref}

\begin{document}

\title{The cyclotron line energy in Her X-1: stable after the decay}

\author{R.~Staubert\inst{1}, L.~Ducci\inst{1}, L.~Ji\inst{1}, F.~F\"urst\inst{2}, 
J.~Wilms\inst{3},  R.E.~Rothschild\inst{4}, K.~Pottschmidt\inst{5}, M.~Brumback \inst{6}, 
F.~Harrison\inst{7} }

\offprints{staubert@astro.uni-tuebingen.de}

\institute{
        Institut f\"ur Astronomie und Astrophysik, Universit\"at T\"ubingen,
        Sand 1, 72076 T\"ubingen, Germany
\and
        European Space Agency - European Space Astronomy Center (ESA-ESAC), Operations Dpt.,
        Camino Bajo del Castillo, s/n., Urb.\ Villafranca del Castillo, 28692 Villanueva de la Canada, 
        Madrid, Spain 
\and
        Dr.\ Remeis Sternwarte \& Erlangen Center for Astroparticle Physics, Univ. Erlangen-N\"urnberg,
        Sternwartstr.~7, 96049 Bamberg, Germany
\and
        Center for Astrophysics and Space Sciences, University of
        California at San Diego, La Jolla, CA 92093-0424, USA
\and
        NASA-Goddard Spaceflight Center, 8800 Greenbelt Rd., Greenbelt, MD 20771, USA,
        Department of Physics and Center for Space Science and Technology, University of Maryland Baltimore County, 
        Baltimore, MD 21250, USA
\and
         Dartmouth College, Department of Physics \& Astronomy, 6127 Wilder Laboratory, Hannover, NH 03755, USA      
\and
        Cahill Center for Astronomy and Astrophysics, California Institute of Technology, 
        Pasadena, CA 91125, USA  
        }

\date{submitted: 06/07/2020, accepted: 17/08/2020 }
\authorrunning{Staubert et al.}
\titlerunning{Constant cyclotron line energy in Her X-1}

\abstract
   {We summarize the results of a dedicated effort between 2012 and 2019 to follow the evolution 
   of the cyclotron line in Her~X-1 through repeated \textsl{NuSTAR}
   observations.
   The previously observed nearly 20-year long decay of the cyclotron line energy has ended around 2012:  
   from there onward the pulse phase averaged flux corrected cyclotron line energy has remained stable
   and constant at an average value of $E_\mathrm{cyc}$ = $(37.44\pm0.07)$\,keV (normalized to a flux
   level of 6.8 \textsl{RXTE}/ASM-cts/s).
   The flux dependence of  $E_\mathrm{cyc}$ discovered in 2007 is now measured 
   with high precision, giving a slope of $(0.675\pm0.075)$\,keV/(ASM-cts/s), corresponding to an increase
   of 6.5\% of $E_\mathrm{cyc}$ for an increase in flux by a factor of two. We also find that all line parameters
   as well as the continuum parameters show a correlation with X-ray flux.
  While a correlation between $E_\mathrm{cyc}$ and X-ray flux (both positive and negative) is
  now known for several accreting binaries with various suggestions for the underlying physics,
  the phenomenon of a long-term decay has so far only been seen in Her~X-1 and Vela~X-1,
  with far less convincing explanations. }

   
   \maketitle

\section{Introduction}

The eclipsing binary Her~X-1/HZ~Her is a low mass X-ray binary (LMXB), 
discovered as an X-ray source by the first X-ray satellite \textsl{UHURU} in 1971
\citep{Tananbaum_etal72}.   
Similar to Cen~X-3, the source was identified as an
X-ray pulsar, powered by mass accretion from its companion. Her~X-1 is one of the
most interesting X-ray pulsars due to its wide variety of observable features.
Of the many introductions to this source we refer to some of the most recent ones,
e.g., \citet{Staubert_etal17, Staubert_etal19, Sazonov_etal20}. In order to maintain
some degree of completeness within this contribution we list the following main features
of Her~X-1: the spin period of the neutron star is 1.24\,s, the orbital period is  1.7\,d 
(identified by eclipses and the modulation of the pulse arrival times), there is a 
super-orbital flux modulation with a somewhat variable period of $\sim$35\,d. 
This \textsl{On-Off} variation can be understood as being due to the precession of 
a warped accretion disk \citep{Petterson_77,SchandlMeyer_94}. Due to the high 
inclination of the binary ($i>80^\circ$) we see the disk nearly edge-on
\citep{GerendBoynton_76}. The precessing 
warped disk covers the central X-ray source during a substantial portion of the 35\,d 
period \citep{Klochkov_etal06,Klochkov_etal08}. 

The X-ray spectrum Her~X-1 is a power law continuum with exponential cutoff,
as typical of accreting binary pulsars \citep{Wolff_etal16}. The cyclotron line around 
37\,keV, discovered in a balloon observation in 1975 \citep{Truemper_etal78}, is 
due to resonant scattering of photons by electrons on Landau levels in the 
$\sim$$10^{12}$\,Gauss magnetic field at the polar caps of the neutron star. 
It is therefore often referred to as  a cyclotron resonant scattering feature (CRSF). 
The energy spacing between the Landau levels is approximately given by 
$E_\mathrm{cyc} \approx 11.6\,\text{keV}\,B_{12}$, 
where $B_{12}$ is the magnetic field strength in units of $10^{12}\,\text{Gauss}$. 
If the gravitational redshift is taken into account, 
the magnetic field strength at the site of the emission of the X-ray spectrum
can be measured directly from the observed energy of the fundamental cyclotron 
line in the X-ray spectrum: 
$B_{12}\approx (1+z)~E_\text{obs}/11.6\,{\rm keV}$, where $z$ is the gravitation
redshift  \citep{Schwarm_etal17a}. 

The discovery of the cyclotron feature in the spectrum of Her X-1 was
the first direct measurement of the surface magnetic field strength of a neutron star. 
Contrary to other ways to estimate such a magnetic field strength, no further model
assumptions are needed. We now know about 35 binary X-ray pulsars that show 
cyclotron lines in their spectra, generally between a few keV and $\sim$100\,keV
(for reviews, see 
\citealt{Staubert_etal19,RevnivtsevMereghetti_16,CaballeroWilms_12,
Wilms_12,Terada_etal07,Heindl_etal04,Staubert_03,Coburn_etal02}).

\begin{table*}
\caption{Details on \textsl{NuSTAR} observations of Her~X-1 between 2012 and 2020. }
\vspace{-3mm}
\begin{center}
\begin{tabular}{lllllllll}
\hline\noalign{\smallskip}
Observation    & Obs ID            & 35-day       & Start          & End           & Center        & Net           & 35-day                     & 35-day    \\
date                &                         & cycle         & of obs        & of obs        & of obs        & expo   & \textsl{Turn-On}$^{b}$   & phase$^{c}$       \\
                       &                         & no.$^{a}$  &                   &                   &                   & sure          &                                 & of center    \\
                       &                         &                  & [MJD]         & [MJD]        & [MJD]         & [ksec]        & [MJD]                      & of obs        \\
\hline\noalign{\smallskip}
22 Sep 2012  & 30002006005$^{d}$ & 427   & 56192.19   & 56192.77   & 56192.48    & $\sim$22  & $56189.0\pm0.1$    & 0.100       \\
03 Aug 2015  & 90102002002   & 457           & 57237.69   & 57238.26   & 57237.98    & 22.5          & $57233.5\pm0.1$    & 0.128        \\
20 Aug 2016  & 10202002002   & 468           & 57620.19   & 57621.26   & 57620.73    & 36.6          & $57617.2\pm0.1$    & 0.101        \\
05 Aug 2017  & 30302012002   & 478           & 57970.42   & 57971.21   & 57970.81    & 28.4          & $57965.7\pm0.2$    & 0.147        \\
26 Feb 2018  & 30302012004   & 484           & 58175.07   & 58175.79   & 58175.43    & 18.3          & $58171.5\pm0.5$    & 0.113        \\
17 Sep 2018  & 30402009002   & 490           & 58378.83   & 58379.56   & 58379.19    & 28.4          & $58377.7\pm0.3$    & 0.044        \\
09 Feb 2019  & 30402034002   & 494           & 58523.41   & 58523.85   & 58523.63    & 18.3          & $58516.6\pm0.2$    & 0.202$^{e}$    \\
14 Mar 2019  & 30402034008   & 495           & 58556.28   & 58556.75   & 58556.51    &   4.3          & $58551.5\pm0.7$    & 0.144        \\
23 Jun 2019  & 30402009004   & 498           & 58657.34   &  58658.06  & 58657.70    & 27.1          & $58654.1\pm0.2$    & 0.102        \\
\noalign{\smallskip}\hline
\end{tabular}
\end{center}
$^{a}$ 35-day cycle numbering is according to \citet{Staubert_etal83}; 
$^{b}$ as determined from the monitoring data of \textsl{Swift}/BAT; \\
$^{c}$ using $P_{35}$ = 34.85\,d; $^{d}$ see \citet{Fuerst_etal13}, Table~1;
$^{e}$ this observation is at a particular high 35-day phase. 
\label{tab:obs}
\end{table*}

\begin{table*}
\caption{Summary of the spectral analysis of nine \textsl{NuSTAR} observations of Her~X-1. The spectral parameters
  were found by applying the XSPEC-function \texttt{highecut} (see text). Uncertainties are at the 1 sigma (68\%) level.
  The maximum flux of the respective 35-day cycle is given in units of (ASM-cts/s), referring to the All Sky Monitor of 
  \textsl{RXTE}. The corresponding physical flux in units of (keV/cm$^{2}$~s) results by multiplying with 0.2367.
  The flux was actually measured by \textsl{Swift}/BAT and converted according to
  (2--10\,keV) ($\mathrm{ASM-cts/s}$) = 93.0~$\times$~(15--50\,$\mathrm{keV}$)~($\mathrm{BAT-cts/cm^{2}}$~s)
\citep{Staubert_etal16}. The observed line energy was normalized to an ASM-count rate of 6.8\,cts/s 
 by using a slope of $(0.675\pm0.075)$\,keV/(ASM-cts/s) (see Fig.~\ref{fig:Ec_flux}). }
\vspace{-6mm}
\begin{center}
\begin{tabular}{llllllllll}
\hline\noalign{\smallskip}
 35\,d & max. flux            & Observed           & Line                   & Line                   & $E_\mathrm{cyc}$         & $E_\mathrm{cut}$ & $E_\mathrm{fold}$ & Power \\
 cycle & of 35\,d              & line                     & width                  & strength$^{a}$  & norm. to 6.8                   &                               &                              & law      \\
 no.$^{e}$  & cycle          & energy                & $\sigma$           &                          & ASM-cts/s                      &                               &                              & index  \\
          & [ASM-cts/s]        & [keV]                  & [keV]                  & [keV]                 & [keV]                              & [keV]                      &  [keV]                    & $\Gamma$ \\
\hline\noalign{\smallskip}
 427  & $6.60\pm0.37$   & $37.40\pm0.25$$^{b}$ & $5.76\pm0.29$ & $8.86\pm0.87$ & $37.54\pm0.25$    & $20.68\pm0.27$   & $9.95\pm0.13$     & $0.920\pm0,004$ \\
 457  & $2.96\pm0.20$   & $34.79\pm0.22$ & $4.46\pm0.22$ & $4.70\pm0.70$ & $37.38\pm0.24$$^{c}$    & $19.86\pm0.12$   & $9.37\pm0.09$     & $0.929\pm0.003$ \\
 468  & $6.50\pm0.20$   & $37.18\pm0.14$ & $5.97\pm0.18$ & $8.83\pm0.44$ & $37.38\pm0.14$$^{c}$    & $20.86\pm0.15$   & $10.16\pm0.07$   & $0.985\pm0.001$ \\
 478  & $4.10\pm0.20$   & $35.62\pm0.18$ & $4.94\pm0.20$ & $5.90\pm0.40$ & $37.44\pm0.19$              & $19.98\pm0.16$   & $9.79\pm0.09$     & $0.962\pm0.002$ \\
 484  & $4.09\pm0.19$   & $35.67\pm0.29$ & $4.84\pm0.33$ & $6.10\pm0.70$ & $37.50\pm0.30$              & $20.04\pm0.11$   & $10.16\pm0.07$   & $0.963\pm0.002$ \\
 490  & $5.60\pm0.46$   & $36.65\pm0.16$ & $5.61\pm0.25$ & $8.44\pm0.59$ & $37.46\pm0.16$              & $20.45\pm0.23$   & $9.79\pm0.09$     & $0.974\pm0.002$ \\
 494  & $5.02\pm0.46$   & $36.28\pm0.22$ & $5.26\pm0.24$ & $7.21\pm0.53$ & $37.48\pm0.23$              & $19.56\pm0.12$   & $9.65\pm0.11$     & $0.885\pm0.002$$^{f}$ \\
 495  & $3.72\pm0.56$   & $35.36\pm0.41$ & $4.76\pm0.45$ & $6.38\pm0.98$ & $37.44\pm0.43$              & $19.62\pm0.29$   & $9.49\pm0.24$     & $0.934\pm0.005$ \\
 498$^{d}$  & $4.00\pm0.37$ & $35.65\pm0.21$ & $5.01\pm0.25$ & $7.02\pm0.50$ & $37.54\pm0.22$      & $19.81\pm0.16$   & $9.38\pm0.09$     & $0.932\pm0.001$ \\
\noalign{\smallskip}\hline
\end{tabular}
\end{center}
$^{a}$ we note that strength = $\sigma~\tau~\sqrt{2\pi}$; \\
$^{b}$ the values for the CRSF are from \citet{Fuerst_etal13}, Obs. II (Table~3, HighE); \\
$^{c}$ in \citet{Staubert_etal16,Staubert_etal17} the flux normalization was done with a slope of 0.44 (instead of 0.675); \\
$^{d}$ for the June 2019 observation only data from detector B have been used (see text); \\
$^{e}$ 35-day cycle numbering is according to \citet{Staubert_etal83}; \\
$^{f}$ observation at a high 35-day phase, $\Gamma$ expected to be lower \citep{Vasco_12}.
\label{tab:results}
\end{table*}

Significant variability has been observed with the CRSF in Her~X-1,
regarding its centroid energy $E_\mathrm{cyc}$ and other characteritic
parameters like its width and its optical depth. These parameters do vary
with pulse phase, with luminosity, and with time \citep{Staubert_etal14}. 
Her~X-1 was in fact the source in which all these variations were 
observed for the first time: a positive correlation between $E_\mathrm{cyc}$ and 
the X-ray luminosity $L_{x}$ \citep{Staubert_etal07} (confirmed on short 
timescales by the pulse-amplitude-resolved technique by \citealt{Klochkov_etal11}), 
and a long-term decay of $E_\mathrm{cyc}$, co-existing with the 
luminosity dependence \citep{Staubert_etal14,Staubert_etal16}. The long-term 
decay was confirmed by \citet{Klochkov_etal15} using monitoring data of
\textsl{Swift}/BAT\footnote{BAT refers to the Burst Alert Telescope on the NASA mission \textsl{Swift}}. 
For the current knowledge about such variations in other accreting X-ray pulsars see \citet{Staubert_etal19}. 

Of particular interest has been the long-term decay and the question whether 
this would end at some time - or even invert, such that $E_\mathrm{cyc}$ would
rise again. This seemed to have been observed in 2017 \citep{Staubert_etal17}.
We do, however, show here that the decay had ended, but a turn-up did not
actually materialize (see Sect. \ref{sec:line_param}).
The end of the decay is supported through observations with \textsl{Swift}/BAT
\citep{Ji_etal19} and \textsl{Astrosat} \citep{Bala_etal20}. 

Here we summarize the results of nine observations of Her~X-1 by
\textsl{NuSTAR}\footnote{\textsl{NuSTAR} refers to the NASA mission Nuclear Spectroscopic Telescope Array}
 in the time frame 2012 to 2019 with regard to the 
cyclotron resonance scattering feature in the pulse averaged X-ray spectrum: 
the CRSF energy has apparently stopped its ${\sim}20$\,-year long decay and has stayed
constant since around 2012. In addition to the CRSF centroid energy also its width and strength 
are clearly correlated with flux, the dependencies are now measured with high precision.
We further present evidence for a dependence of all continuum parameters on X-ray flux.

\section{Observations and analysis}

In Table~\ref{tab:obs} we list nine observations of Her~X-1 performed by \textsl{NuSTAR} 
\citep{Harrison_etal13} over the time period 2012 to 2019, all  done close to the maximum flux 
of a \textsl{Main-On} state. Also given are the net exposure times (varying between 4.3\,ks and 
36.6\,ks), the times of the respective 35-day \textsl{Turn-On} and the 35-day phase of the center of the
respective observations.
The details of the data analysis are similar to those described by \citet{Staubert_etal14} and \citet{Staubert_etal16}. 
We used the standard \texttt{nupipeline} and \texttt{nuproducts} utilities (01 Apr 20\_v1.9.2) and
XSPEC\footnote{https://heasarc.gsfc.nasa.gov/xanadu/xspec/} v12.11 as
part of HEASOFT\footnote{http://heasarc.nasa.gov/lheasoft, 6.27.2, caldb release 20191219.}.
The source extraction diameter was selected between 90 arcsec and 120 arcsec depending on the
brightness of the source. All values given are from  simultaneous spectral fitting of the data from 
both focal plane detectors, unless otherwise stated. The spectral function used for all observations 
was the XSPEC-function \texttt{highecut} in combination with a \texttt{power law}:
\begin{equation}
I_E =
\begin{cases}
K\cdot E^{-\Gamma}, & \text{if\,} E \leq E_\text{cut} \\
K\cdot E^{-\Gamma} \exp{\left(-\frac{E-E_\text{cut}}{E_\text{fold}}\right)}, & \text{if\,}
E > E_\text{cut}.
\end{cases}
\end{equation}
where $\Gamma$ is the power law (photon) index, $E_\text{cut}$ is the energy where the
cut-off sets in, and $E_\text{fold}$ is the e-folding energy describing the flux decay.
The function contains a discontinuity of its first derivative (a ``break'') at $E=E_\text{cut}$. 
In order to smooth this break, generally an artificial small Gaussian absorption line was added 
with the center energy fixed to $E_\text{cut}$ and a free width and depth. Neither the power law, nor
the exponential cut off are affected by this \citep[e.g.,][]{Coburn_etal02}.

The cyclotron line is modeled by the Gaussian shaped "\texttt{gabs}" function: To model the 
cyclotron line, one modifies the continuum functions described above by the inclusion of a 
corresponding multiplicative component of the form $e^{-\tau(E)}$, where the optical depth 
$\tau(E)$ has a Gaussian profile:
\begin{equation}
\tau(E) = \tau_0\exp\left[-\frac{(E-E_\text{cyc})^2}{2\sigma_\text{cyc}^2}\right],
\end{equation}
with $\tau_0$, $E_\text{cyc}$, and $\sigma_\text{cyc}$ 
being the central optical depth, the centroid energy, and the width of the line. 
We note that in the popular \textsl{XSPEC} realization of this function
\texttt{gabs}, $\tau_0$ is not explicitely used as a fit parameter. Instead, a product 
$\tau_0\sqrt{2\pi}\sigma_\text{cyc}$ 
is defined as the \textsl{``strength''} of the line.

\begin{figure}
\includegraphics[angle=90,width=11.1cm]{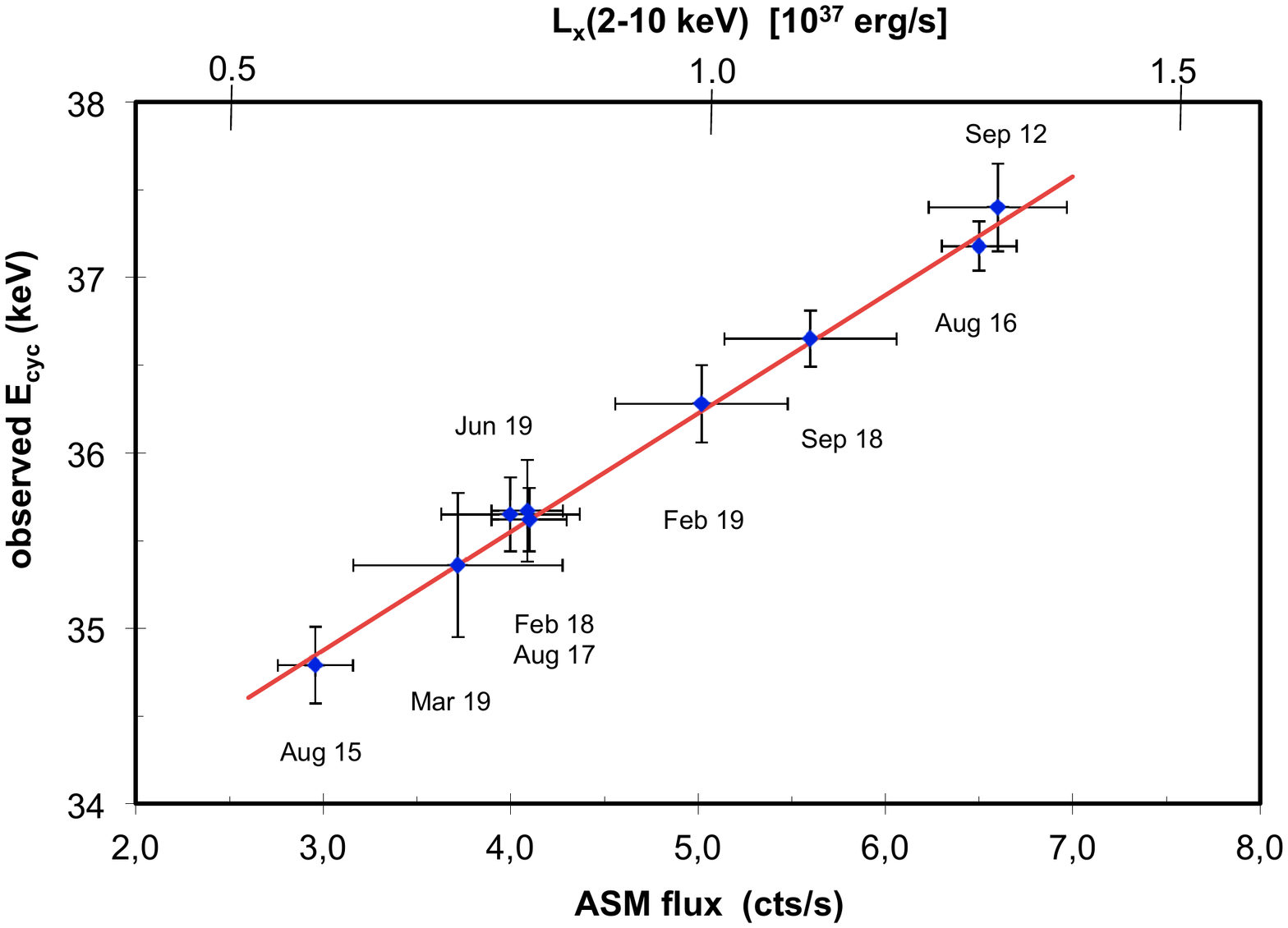}
\vspace*{-14mm}
\caption{Correlation between the measured values of the pulse phase averaged 
 cyclotron line energy and the X-ray flux (at the maximum of the respective 35d Main-On),
 as measured by \textsl{Swift}/BAT (in units of (ASM-cts/s))
 for all \textsl{NuSTAR} observations between 2012 and March 2019 (see Table~\ref{tab:obs}).
 We note that 1 (ASM-cts/s) equals 93.0~(Swift-BAT-cts)/(cm$^{2}$~s) \citep{Staubert_etal16}
 and $\mathrm{0.224\,(keV/cm^{2}~s)}$ in (2--10\,keV).
 The best fit line defines a slope of $\mathrm{(0.675\pm0.075)\,keV/(ASM-cts/s)}$.
 Pearsons linear correlation coefficient is 0.98. For the mean power law index of -0.953
 and an adopted distance of 6.6\,kpc \citep{Reynolds_etal97},  the Her~X-1 luminosity is 
 L$_{x}$(2-10\,keV) [$10^{37}$~erg/s]~=~0.187(1)~$\times$~(ASM-cts/s)}.
\label{fig:Ec_flux}
\end{figure}

Some of the listed \textsl{NuSTAR} observations have been performed in
coordination with other satellites, like \textsl{INTEGRAL}, \textsl{Insight}-HXMT
and \textsl{Astrosat}\footnote{\textsl{INTEGRAL} is the International Gamma-ray 
Astrophysics Laboratory of ESA, \textsl{Insight}-HXMT the Chinese mission Hard X-ray 
Modulation Telescope, and \textsl{Astrosat} the X-ray satellite mission of India}
to study the CRSF. The February and March 2019 observations
were coordinated with \textsl{XMM}-Newton\footnote{\textsl{XMM}-Newton is ESAs
Multi Mirror soft X-ray mission} 
for a different project\footnote{Brumback et al. 2020,
submitted}, but also gave data on the CRSF.
Here we will not report on the results from the other
satellites because we are still working on trying to resolve some inconsistencies,
which are likely due to imperfect inter-calibration between the different 
instruments and possibly aging of some of them\footnote{We are still attempting
to perform further simultaneous observations between \textsl{NuSTAR},
\textsl{INTEGRAL} and \textsl{Insight}-HXMT, some are already planned.} -
we plan to report about this in a forthcoming paper.
For first results from \textsl{Insight}-HXMT and \textsl{Astrosat} see \citet{Xiao_etal19} 
and \citet{Bala_etal20}.

\begin{figure*}
\includegraphics[angle=90,width=18cm]{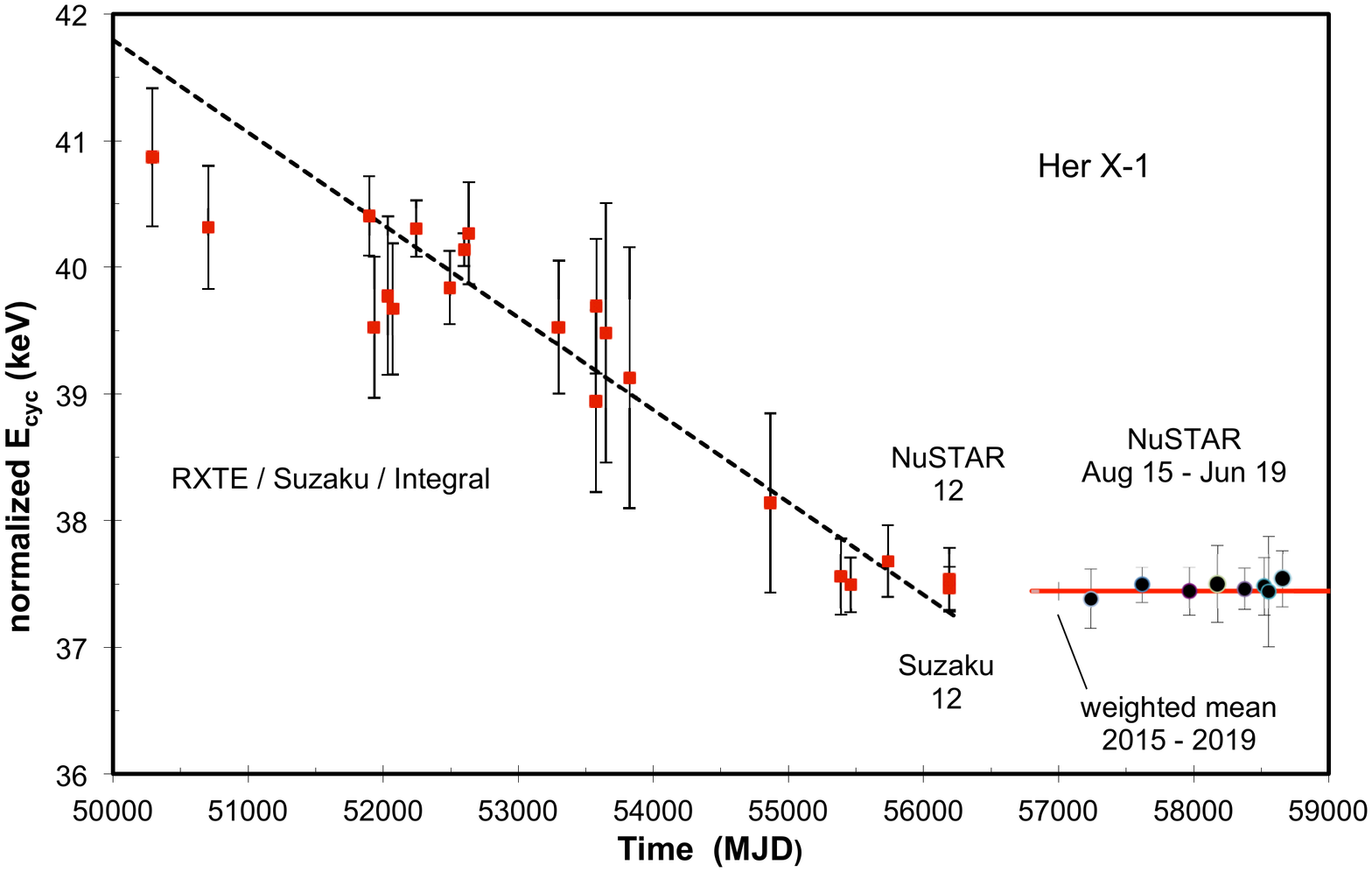}
\vspace*{-13mm}
\caption{Evolution of the cyclotron line energy $E_\mathrm{cyc}$ in Her~X-1.  The red 
  points until 2012 (MJD 56200) and the corresponding linear best fit (dashed line) are 
  reproduced from \citet{Staubert_etal16}. 
  The black points are the new measurements by \textsl{NuSTAR} from 2015-2019
  (see Table~\ref{tab:obs}): the pulse phase averaged $E_\mathrm{cyc}$ values
  normalized to a reference ASM count rate of $\mathrm{6.8\,(ASM-cts/s)}$
  using a flux dependence of $\mathrm{0.675\,keV/(ASM-cts/s)}$
 (see Fig.~\ref{fig:Ec_flux}). The solid red line represents the weighted mean
  of $(37.44\pm0.07)$\,keV, demonstrating a constant value since at least 2012.}
\label{fig:Ecyc_norm}
\end{figure*}

\begin{figure}
\includegraphics[angle=90,width=10.9cm]{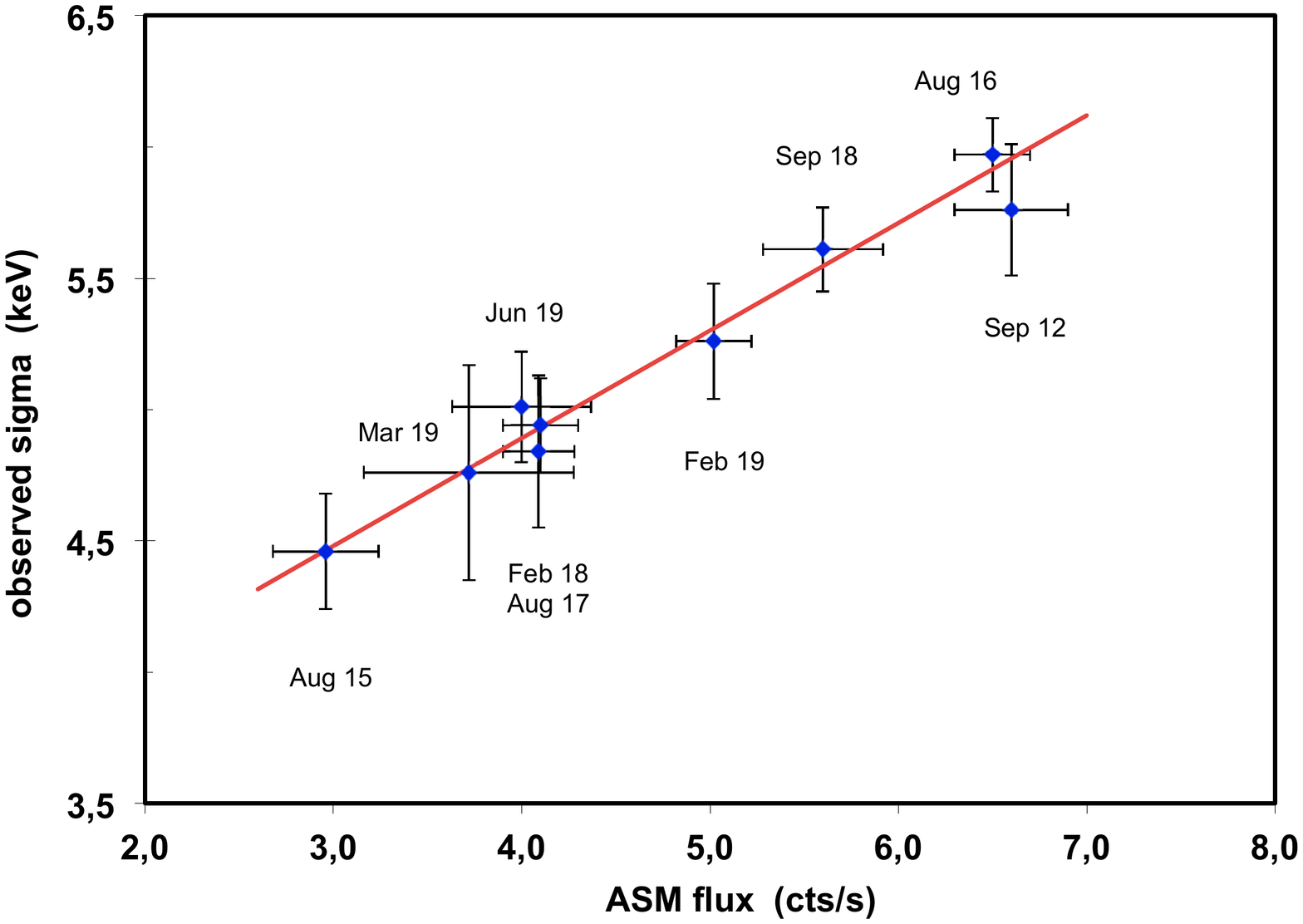}

\vspace*{-13mm}
\caption{Correlation between the measured values of the width (sigma) of the 
 cyclotron line and the X-ray flux. The fit takes into account 
 nine measurements by \textsl{NuSTAR} from 2012-2019  (see Table~\ref{tab:obs}). 
 The best fit line is given by the function
$sig_\mathrm{cyc}$ = $\mathrm{(5.30\pm0.09) + (0.41\pm0.07) \times ((ASM -cts/s)- 5.0)}$
(all values in keV). Pearsons linear correlation coefficient is 0.99.}
\label{fig:sig_ASM}
\end{figure}

\begin{figure}
\includegraphics[angle=90,width=10.9cm]{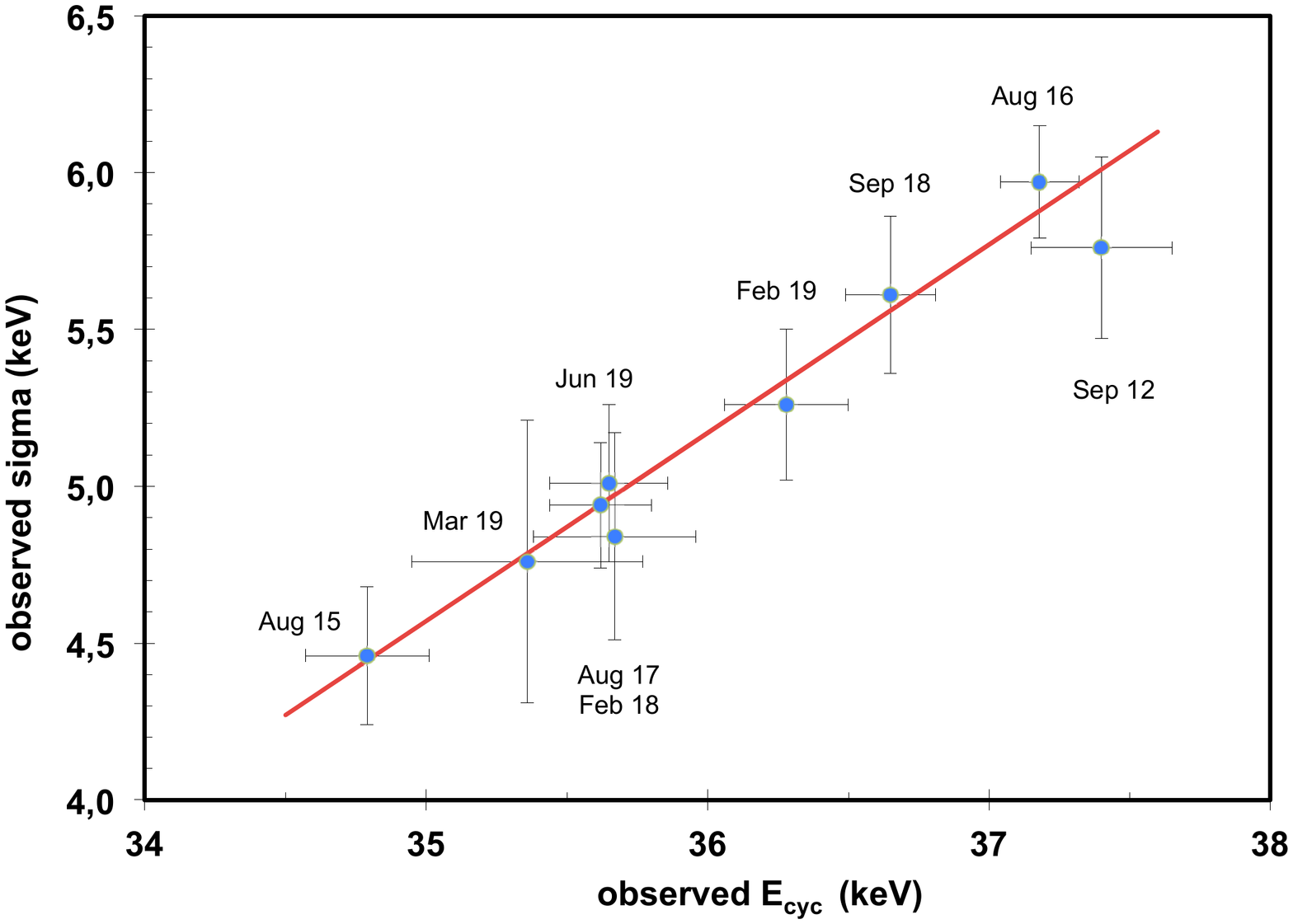}

\vspace*{-13mm}
\caption{Correlation between the width (sigma) of the pulse phase averaged 
 cyclotron line and its centroid energy. The fit takes into account 
 nine measurements by \textsl{NuSTAR} from 2012-2019  (see Table~\ref{tab:obs}).
 The best fit line is given by the function
$\sigma_\mathrm{cyc}$ = $(5.17\pm0.09)+(0.60\pm0.11) \times (E_\mathrm{cyc} - 36.0)$
(all values in keV). Pearsons linear correlation coefficient is 0.96.}
\label{fig:sigcyc_Ecyc}
\end{figure}

\section{Results}

In Table~\ref{tab:results} we summarize the results of the spectral analysis, both for
the cyclotron line (the centroid energy, the width and the strength) and for the continuum 
(cut-off energy $E_\mathrm{cut}$, e-folding energy $E_\mathrm{fold}$ and power law index $\Gamma$).
We further list the maximum fluxes of the respective 35-day cycles. 

In order to allow us to do a comparison to previous results, we use the
observational flux units of (ASM-cts/s), referring to the All Sky Monitor of 
\textsl{RXTE})\footnote{\textsl{RXTE} refers to the NASA mission Rossi X-ray Timing Explorer, and 
ASM to the All Sky Monitor on this satellite},
using the conversion (2--10\,keV) (ASM-cts/s) = 93.0~$\times$~(15--50\,keV)~(BAT-cts/cm$^{2}$~s).
This was found by \citet{Staubert_etal16} by comparing flux values measured by the All Sky Monitor onboard
of \textsl{RXTE} on the one hand and those from \textsl{Swift}/BAT on the other, for the  overlapping time 
of both missions. 

The relationship between ASM (or BAT) count rates and \textsl{NuSTAR} and corresponding
physical flux units was established in the following way: the observed maximum ASM count rate
(from the monitoring observations by \textsl{Swift}/BAT) for each 35d cycle (Table~\ref{tab:results})
was plotted against the normalization, the flux at 1\,keV, as determined
through the spectral analysis of the corresponding \textsl{NuSTAR} observation (cycle 494/Feb 2019
was excluded, since this observation was at a 35d-phase of 0.202, after the maximum flux).
This establishes the relationship: flux at 1\,keV [photons/cm$^{2}$~s~keV]~=~0.0255~$\times$~(ASM-cts/s),
or flux at 1\,keV [photons/cm$^{2}$~s~keV]~=~3.371~$\times$~(BAT-cts/cm$^{2}$~s). The energy flux 
was found by integrating the spectrum over the interested energy range. The following relationships in 
physical units emerge (taking the mean power law index of -0.953):
(2-10\,keV flux) [keV/cm$^{2}$~s]~=~0.224~$\times$~(ASM-cts/s), or 
(2-10\,keV flux) [erg/cm$^{2}$~s]~=~3.58~10$^{-10}$~$\times$~(ASM-cts/s). With a distance of 6.6\,kpc 
to Her~X-1 \citep{Reynolds_etal97}, the corresponding (2-10\,keV) luminosities are given by
L$_{x}$(2-10\,keV) [$10^{37}$~erg/s]~=~0.187(1)~$\times$~(ASM-cts/s), or
L$_{x}$(2-10\,keV) [$10^{37}$~erg/s]~=~17.4(1)~$\times$~(BAT-cts/cm$^{2}$~s).
Also listed in Table~\ref{tab:results} are the cyclotron line energies normalized to a flux of 6.8\,(ASM-cts/s)
using the determined linear flux dependence (Fig.~\ref{fig:Ec_flux}):
$E_\mathrm{cyc-norm}$ [keV] = $E_\mathrm{cyc-obs}$ [keV]~+~0.675~$\times$~((ASM-cts/s)~-6.8).
This relation is extremely well established (with a Pearsons linear correlation coefficient of 
0.98\footnote{see, e.g., Numerical Recipes, W.H. Press et al., Cambridge University Press, 1986}). 
The choice of 6.8\,(ASM-cts/s) as reference flux is historical and allows a direct comparison to 
previous results (for all other spectral parameters we now use a reference flux of 5.0\,(ASM-cts/s)
because this flux is closer to the center of the flux range observed, and nearly corresponds to
a (2-10\,keV) luminosity of $\sim10^{37}$~erg/s).

Generally, all parameter values stated are from the combined spectral
analysis of both focal plane detectors of \textsl{NuSTAR}, except for the observation in June 2019 
(cycle 498): only focal plane detector B was used, for $E_\mathrm{cut}$ the value from detector A
is exceptionally high and far outside the overall trend (Fig.~\ref{fig:Ecut_ASM}) (the anomaly is
being investigated).

\begin{figure}
\includegraphics[angle=90,width=10.5cm]{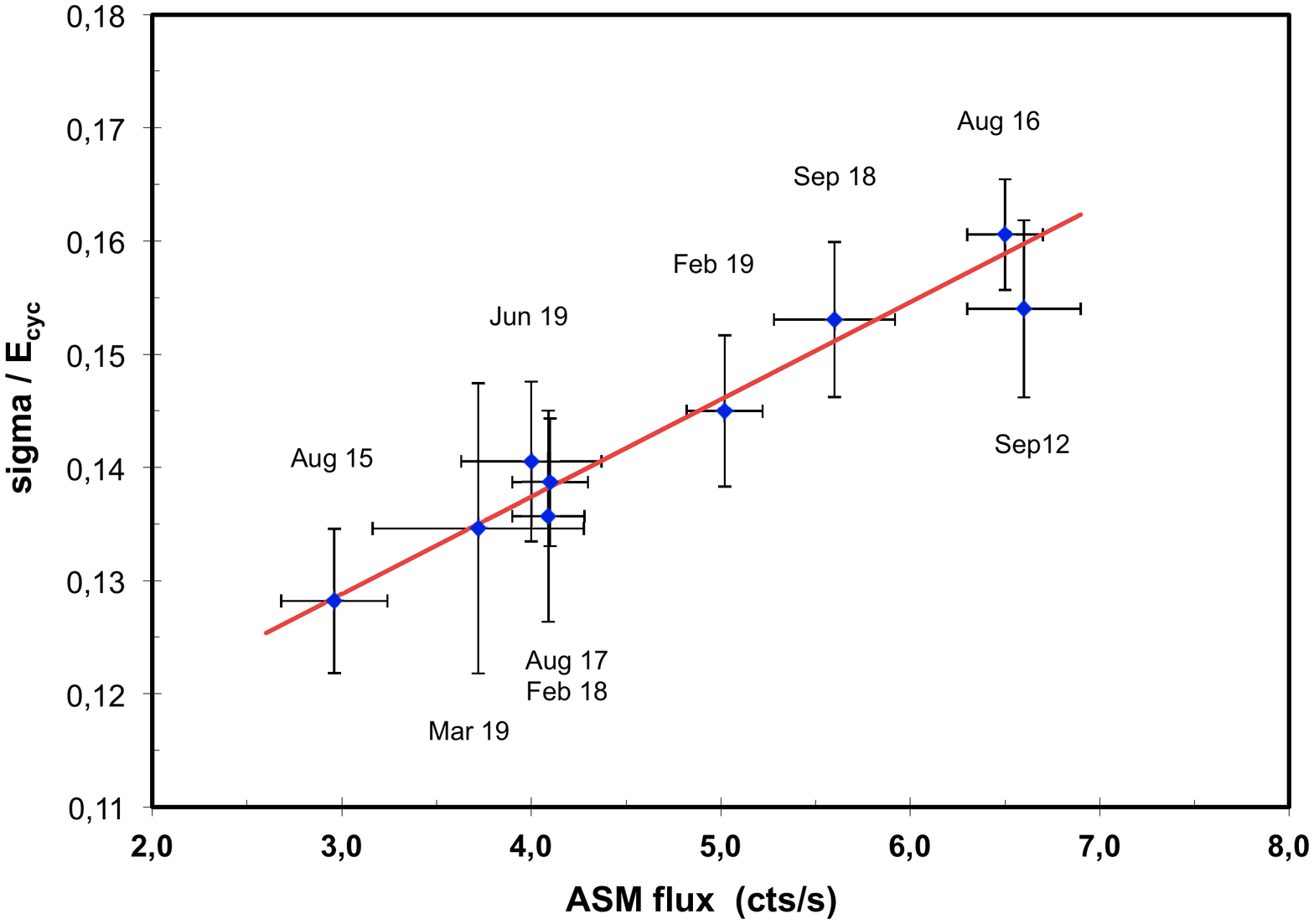}
\vspace*{-13mm}
\caption{The relative width of the cyclotron line versus X-ray flux in units of (ASM-cts/s).
 The best fit line is given by the function
$\sigma_\mathrm{cyc}$ / $E_\mathrm{cyc}$ = $(0.146\pm0.002)+(0.0086\pm0.0019) \times ((ASM-cts/s) - 5.0)$.
Pearsons linear correlation coefficient is 0.97.}
\label{fig:relat_sig_ASM}
\end{figure}

\subsection{Cyclotron line parameters}
\label{sec:line_param}

The new data allow us to determine the correlation between the observed pulse averaged  
cyclotron line energy and the X-ray flux with a significantly improved accuracy when
compared to the time of the discovery of this correlation \citep{Staubert_etal07}.
Fig.~\ref{fig:Ec_flux} shows the definite correlation which can be described by the linear function 
$E_\mathrm{cyc} ~[\mathrm{keV}]$ = $a+b\times~\mathrm{((ASM-cts)/s-6.8)}$,
with $a = (37.44\pm0.07)$ being the CRSF value at an ASM count rate of 6.8\,cts/s  and
$b = \mathrm{(0.675\pm0.075)\,keV/(ASM-cts)/s}$ the slope describing the
flux dependence.

Figure~\ref{fig:Ecyc_norm} displays the evolution of the normalized CRSF centroid
energy of Her X-1 from 2009 to 2019. The red data points are historical
results that were published before, together with the dashed line representing 
the end of the phase of the long-term decay of $E_\mathrm{cyc}$ between
1996 and 2012 as reported by \citet{Staubert_etal16} (their Fig.~2).
The right hand side shows the latest values from \textsl{NuSTAR} (2015-2019).
All data points are flux corrected (normalized to an ASM cout rate of  6.8\,cts/s). 
The values since 2015 are apparently consistent with a constant value, the formal 
weighted average is $\langle E_\mathrm{cyc}\rangle(2015-2019)$ = $(37.44\pm0.07)$\,keV.
Because there is no time dependence of the normalized $E_\mathrm{cyc}$ in 2015--2019, 
it is not necessary to perform a combined fit with simultaneously existing flux and time 
dependencies, as was done for the data earlier than 2012 \citep{Staubert_etal16}.

We need to point out, however, that the new data require a modification of 
the view presented in \citet{Staubert_etal17}, in which it was suggested that an inversion
(an upward trend) in $E_\mathrm{cyc}$ had occured after the end of the decay.
This impression was mainly driven by the 2015 measurement which happened
at an extremely low flux level - in fact the lowest of all \textsl{NuSTAR}
observations at around 3\,(ASM-cts)/s (see Fig.~\ref{fig:Ec_flux}). 
This had, on the one hand, turned out to be useful in extending the dynamic 
range in observed fluxes beyond the (classical) factor of two, it led,
on the other hand, to a very low value of the flux normalized 
$E_\mathrm{cyc}$ when the correction was done with the then best value
of the flux dependence of 0.44\,$\mathrm{keV/(ASM-cts/s)}$
\citep{Staubert_etal16,Staubert_etal17}. When the current best value 
(0.675 instead of 0.44) is used for the normalization, the 2015 value
is significantly higher and consistent with values found throughout 2012--2019 
(see Table~\ref{tab:obs} and Fig.~\ref{fig:Ecyc_norm}).

\begin{figure}
\includegraphics[angle=90,width=10.5cm]{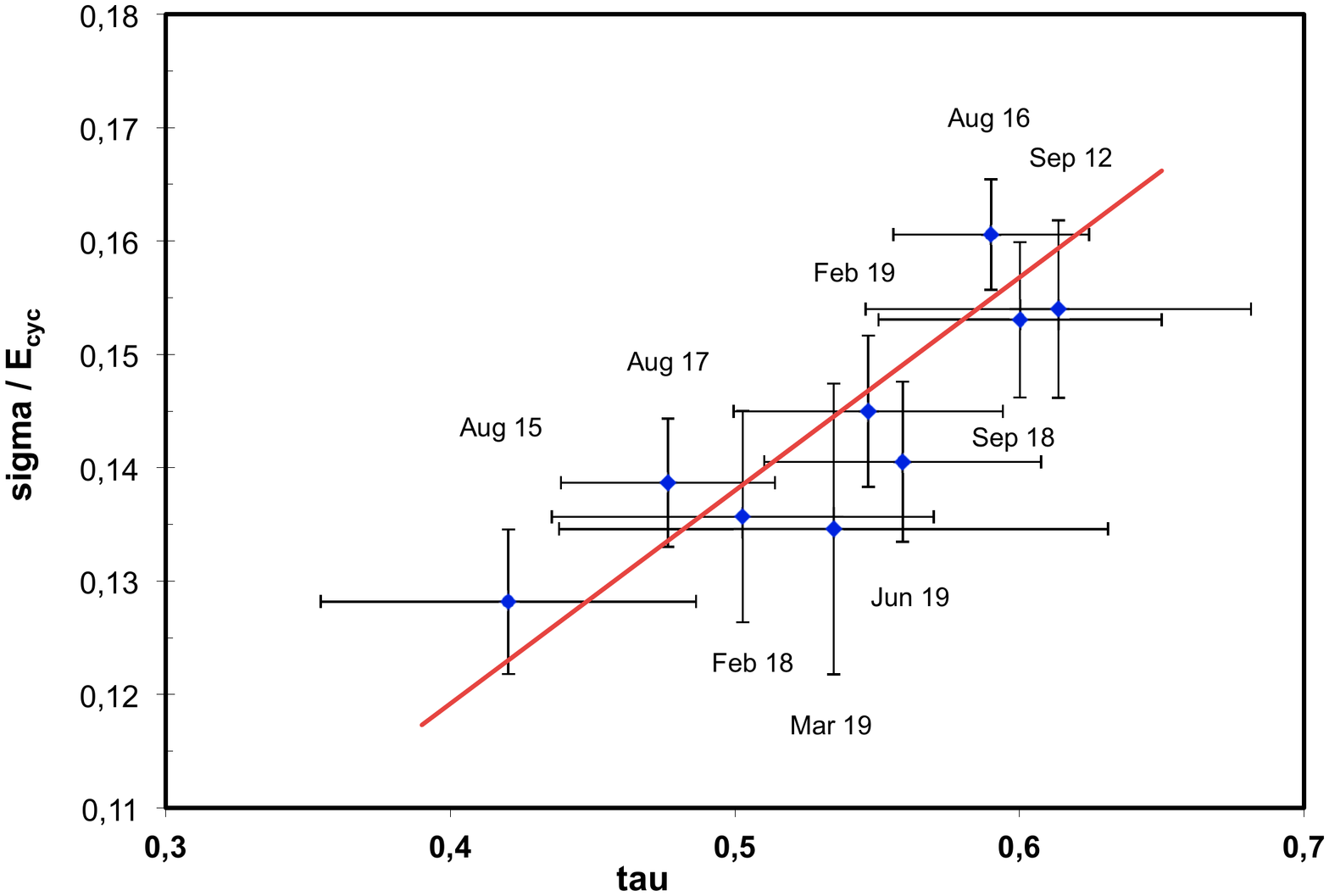}
\vspace*{-13mm}
\caption{The relative width of the cyclotron line versus optical depth $\tau$.
 The best fit line is given by the function
$\sigma_\mathrm{cyc}$ / $E_\mathrm{cyc}$ = $(1.38\pm0.005)+(0.188\pm0.076) \times (\tau_\mathrm{cyc} - 0.5)$.
Pearsons linear correlation coefficient is 0.88.}
\label{fig:relat_sig_tau}
\end{figure}

A further result of our analysis of the nine \textsl{NuSTAR} observations is that we find that
also all other characteristic parameters of the cyclotron line -- the width $\sigma$, the "strength", 
the optical depth $\tau$ and the relative width - are linearly correlated with the X-ray flux. 
In Table~\ref{tab:vs_ASM} we summarize the results of linear fits of all the CRSF 
line parameters versus flux. This also means that all line parameters correlate with one-another 
linearly. As an example, Fig.~\ref{fig:sig_ASM} shows the dependence of the line width 
$\sigma$ on flux and Fig. ~\ref{fig:sigcyc_Ecyc} the dependence of $\sigma$ 
on $E_\mathrm{cyc}$. This correlation is a well known behavior (apparently valid for all cyclotron 
line sources, e.g., \cite{Makishima_etal99,Coburn_etal02}), 
that is expected to occur through thermal Doppler broadening because of the free 
movement of electrons along the magnetic field lines (see discussion below). 
Even though the correlations between the different parameters can in principle be re-constructed 
from the respective dependencies of all parameters on flux (Table~\ref{tab:vs_ASM}), we have 
performed the linear fits for every possible pair of parameters explicitely and summarize the results 
in Table~\ref{tab:inter_relations}.

\begin{table*}
\caption{Relations between line parameters, $y$ = a + b (x - c): parameter y (top line) versus parameter $x$ (first column).
The offset $c$ is constant for any given parameter $x$. Uncertainties are at the 1 sigma (68\%) level.}
\vspace{-3mm}
\begin{center}
\begin{tabular}{llllll}
\hline\noalign{\smallskip}
Parameter               & $E_\mathrm{cyc}$   & $\sigma_\text{cyc}$    & strength                      & $\tau$                            & $\sigma_\text{cyc}$ / $E_\mathrm{cyc}$ \\
 versus                    & [keV]                        & [keV]                           & [keV]                           &                                       &   \\
\hline\noalign{\smallskip} 
 $E_\mathrm{cyc}$   & ----                           &  $a = 5.17\pm0.09$     &  $a = 6.91\pm0.22$   &  $a = 0.53\pm0.02$       &  \\ 
                                 & ----                            &  $b = 0.60\pm0.11$     &  $b = 1.71\pm0.29$   &  $b = 0.064\pm0.023$  &  \\
                                 & ----                            &  $c = 36.0$\,keV          &  $c = 36.0$\,keV        &  $c = 36.0$ \,keV          &  \\                      
\hline\noalign{\smallskip} 
$\sigma_\text{cyc}$  &                                  & ----                               &  $a = 6.42\pm0.31$   &  $a = 0.51\pm0.02$      &  \\     
                                 &                                   & ----                               &  $b = 2.84\pm0.63$   &  $b = 0.11\pm0.04$      &  \\
                                 &                                   & ----                               &  $c = 5.0$\,keV          &  $c = 5.0$\,keV             &   \\ 
\hline\noalign{\smallskip} 
 strength                   &                                   &                                     & ----                             &  $a = 0.53\pm0.02$       &  \\            
                                 &                                   &                                     & ----                             &  $b = 0.041\pm0.014$   &  \\
                                 &                                   &                                     & ----                             &   $c = 7.0$                      &  \\
\hline\noalign{\smallskip} 
 $\tau$                      &                                    &                                    &                                   & ----                                 & $a = 0.14\pm$0.05    \\            
                                 &                                    &                                    &                                   & ----                                 & $b = 0.19\pm$0.08    \\
                                 &                                    &                                    &                                   & ----                                 & $c = 0.50$                   \\
 \hline\noalign{\smallskip} 
 $E_\mathrm{cut}$   & $a = 35.86\pm0.11$    &                                    &                                   &                                      &  \\
                                 & $b = 1.87\pm0.36$      &                                    &                                   &                                      &  \\
                                 & $c = 20.0$\,keV           &                                    &                                   &                                      &  \\
\hline\noalign{\smallskip} 
 $E_\mathrm{fold}$  & $a = 36.69\pm0.13$     &                                    &                                   &                                      &  \\
                                 & $b = 2.71\pm0.39$       &                                    &                                   &                                      &  \\
                                 & $c = 10.0$\,keV            &                                    &                                   &                                      &  \\                           
\noalign{\smallskip}\hline 
\end{tabular}
\end{center}
\label{tab:inter_relations}
\end{table*}

It is worth to note that the relative line width, $\sigma$ / $E_\mathrm{cyc}$, is not constant, but
increases with increasing flux (Fig.~\ref{fig:relat_sig_ASM}) according to
$\sigma$ / $E_\mathrm{cyc}$~=~0.146~+~0.0087~(ASM~-~5.0). This means that the relative 
change with changing flux is stronger for the line width than for the line energy.

In addition, we give the linear correlation between the relative line width $\sigma$ / $E_\mathrm{cyc}$ 
to the optical depth $\tau$ (Fig.\ref{fig:relat_sig_tau}), a correlation first noticed by \citet{Coburn_etal02}
in a group of cyclotron line objects. As with other correlations, also this one can be  realized in individual 
objects  - here Her~X-1, also in 4U~1538$-$52 \citep{Rodes-Roca_etal08}. This may not be so easy to
understand in the context of theoretical considerations (see discussion).

\begin{figure}
\includegraphics[angle=90,width=10.7cm]{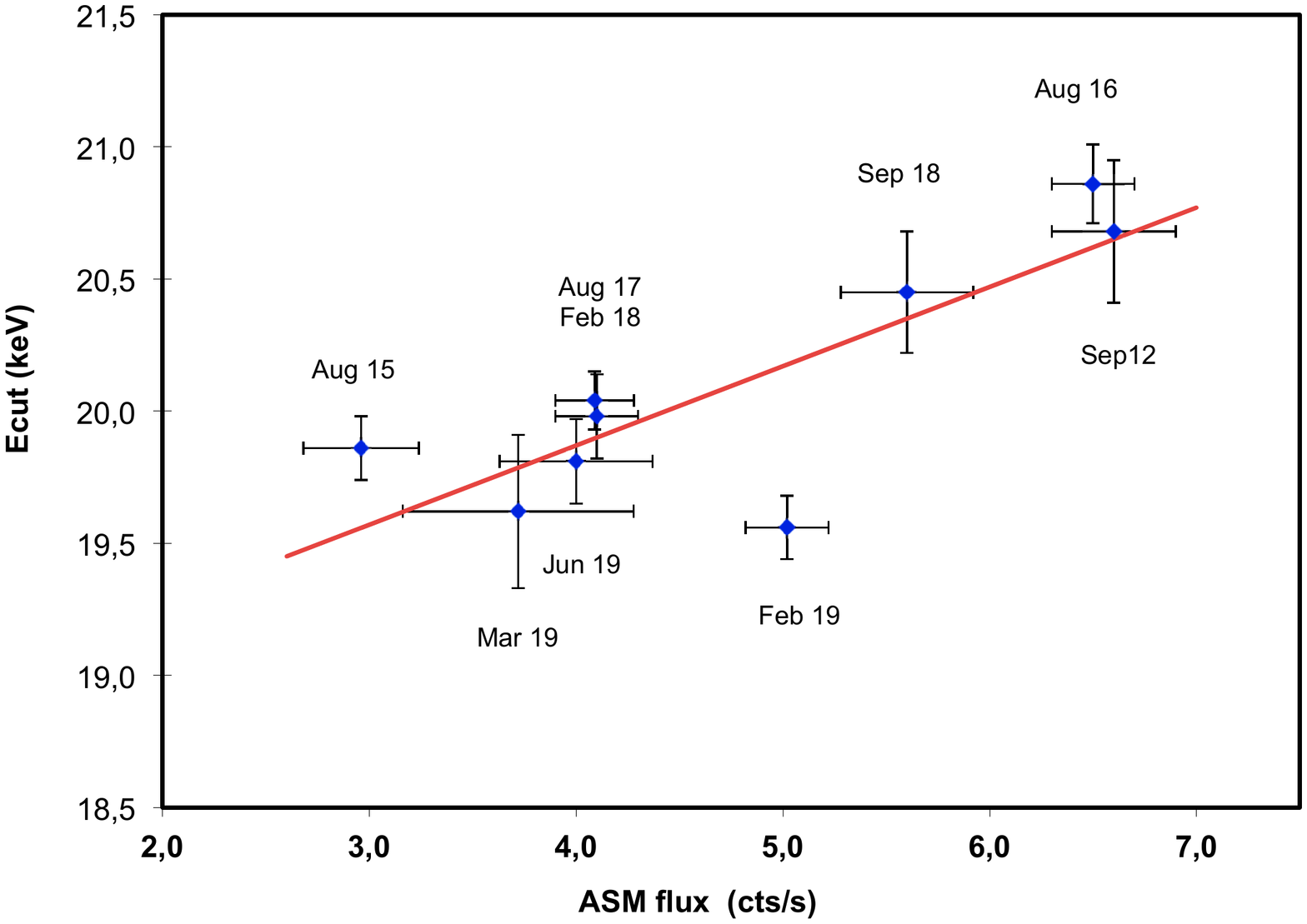}
\vspace*{-13mm}
\caption{The continuum parameter  $E_\mathrm{cut}$ versus X-ray flux in units of (ASM-cts/s).
The value for the June 2019 observation is from focal plane detector B only (see text).
The best fit line is given by the function
$E_\mathrm{cut}$ = $(20.17\pm0.06)+(0.30\pm0.06) \times (ASM - 5.0)$ (all values in keV).
Pearsons linear correlation coefficient is 0.74.}
\label{fig:Ecut_ASM}
\end{figure}

\begin{figure}
\includegraphics[angle=90,width=10.7cm]{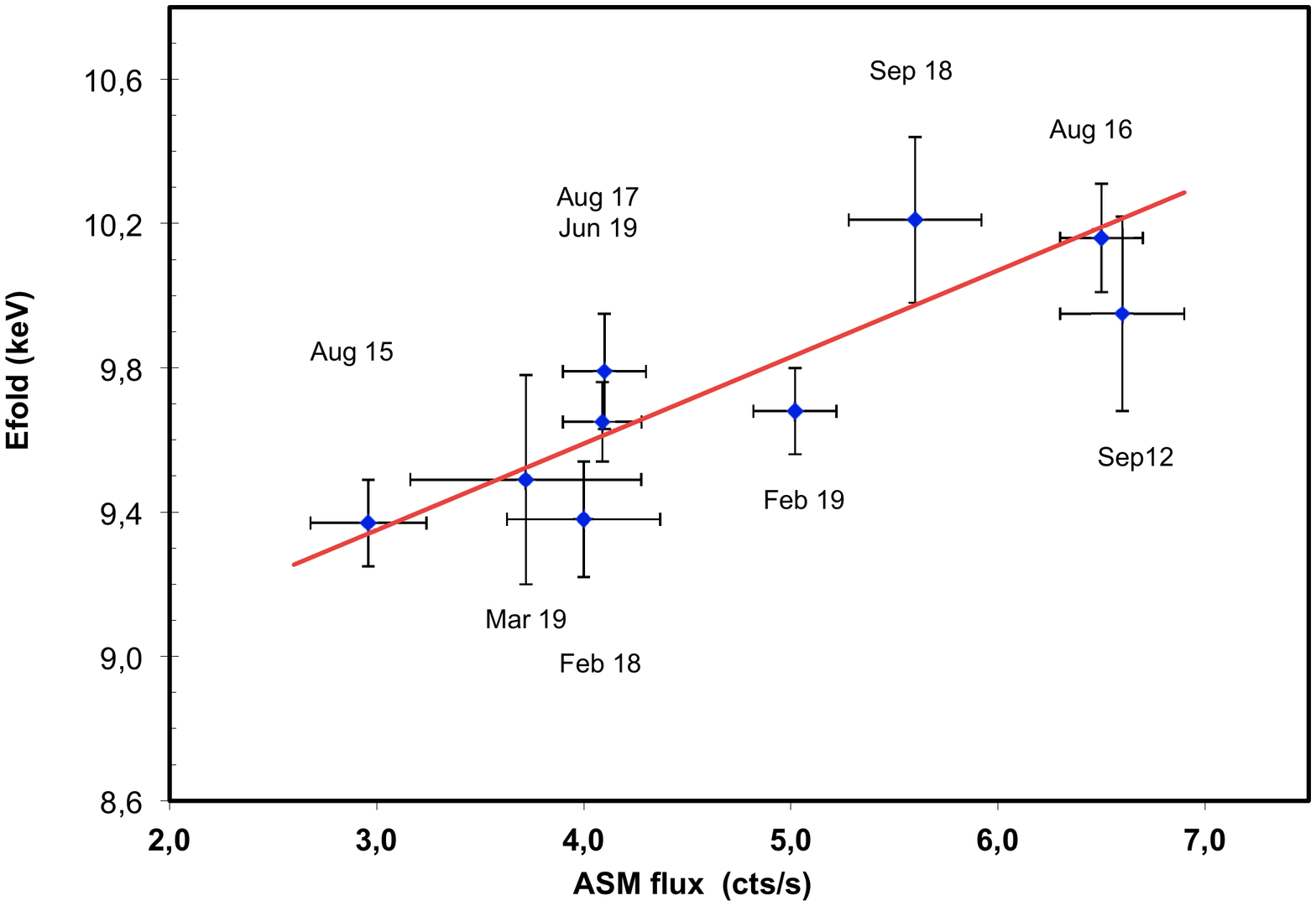}
\vspace*{-13mm}
\caption{The continuum parameter  $E_\mathrm{fold}$ versus X-ray flux in units of (ASM-cts/s).
 The best fit line is given by the function
$E_\mathrm{fold}$ = $(9.83\pm0.04)+(0.24\pm0.03) \times ((ASM-ct/s) - 5.0)$ (all values in keV).
Pearsons linear correlation coefficient is 0.79.}
\label{fig:Efold_ASM}
\end{figure}

\begin{table}
\caption{The linear dependence of spectral parameters on X-ray flux. The five line parameters are
the centroid energy $E_\mathrm{cyc}$, the width $\sigma_\text{cyc}$, the line "strength",
the optical depth $\tau$ (see eq.~2) and the relative width $\sigma_\text{cyc}$ / $E_\mathrm{cyc}$.
The three continuum parameters are $E_\mathrm{cut}$ and $E_\mathrm{fold}$ and the power law
photon index $\Gamma$.
The X-ray flux is measured in units of (ASM-cts/s), referring to the All Sky Monitor of \textsl{RXTE}): 
y = a + b (x - c). Here, the offset in x  is always constant: c = 5.0 (ASM-cts/s) (close to the center of the
range of fluxes observed). We note that  5.0 (ASM-cts/s) corresponds to
5.0$\times$~0.237 = 1.18 (keV/cm$^{2}$~s) in (2--10\,keV). Uncertainties are at the 1 sigma (68\%) level.
The last column lists Pearsons linear correlation coefficients. }
\vspace{-8mm}
\begin{center}
\begin{tabular}{llll}
\hline\noalign{\smallskip}
Parameters                                                &  $a$ [keV] or           & $b$$^{d}$                 & Pearsons\\
                                                                   &  no dimension         &                                  & corr. coeff.\\
\hline\noalign{\smallskip} 
$E_\mathrm{cyc}$$^{a}$ [keV]                  &  $36.24\pm0.09$     & $0.675\pm0.075$     & 0.98   \\ 
$\sigma_\text{cyc}$$^{b}$ [keV]                &  $5.30\pm0.09$       & $0.41\pm0.07$         & 0.99    \\ 
strength$^{c}$                                            & $7.28\pm0.22$        & $1.15\pm0.19$         & 0.96    \\ 
opt. depth $\tau$$^{c}$                              & $0.54\pm0.02$        & $0.042\pm0.015$     & 0.87    \\ 
 $\sigma_\text{cyc}$ / $E_\mathrm{cyc}$  & $0.146\pm0.002$    & $0.0086\pm0.0002$  & 0.97   \\ 
$E_\mathrm{cut}$ [keV]                             & $20.17\pm0.06$      & $0.30\pm0.06$          & 0.74   \\
$E_\mathrm{fold}$ [keV]                            & $9.83\pm0.04$        & $0.24\pm0.03$          & 0.79   \\ 
$\Gamma$                                                 & $0.965\pm0.002$    & $0.015\pm0.001$       & 0.53  \\ 
\noalign{\smallskip}\hline
\end{tabular}
\end{center}
$^{a}$ Fig.~\ref{fig:Ec_flux}; 
$^{b}$ Fig.~\ref{fig:sig_ASM}; 
$^{c}$ we note that strength  [keV] = $\sigma \times \tau \times \sqrt{2\pi}$; \\
$^{d}$ [keV/(ASM-cts/s)] or [1/(ASM-cts/s)]
\label{tab:vs_ASM}
\end{table}

\subsection{Continuum parameters}
\label{sec:cont_param}

\begin{figure}
\includegraphics[angle=90,width=10.7cm]{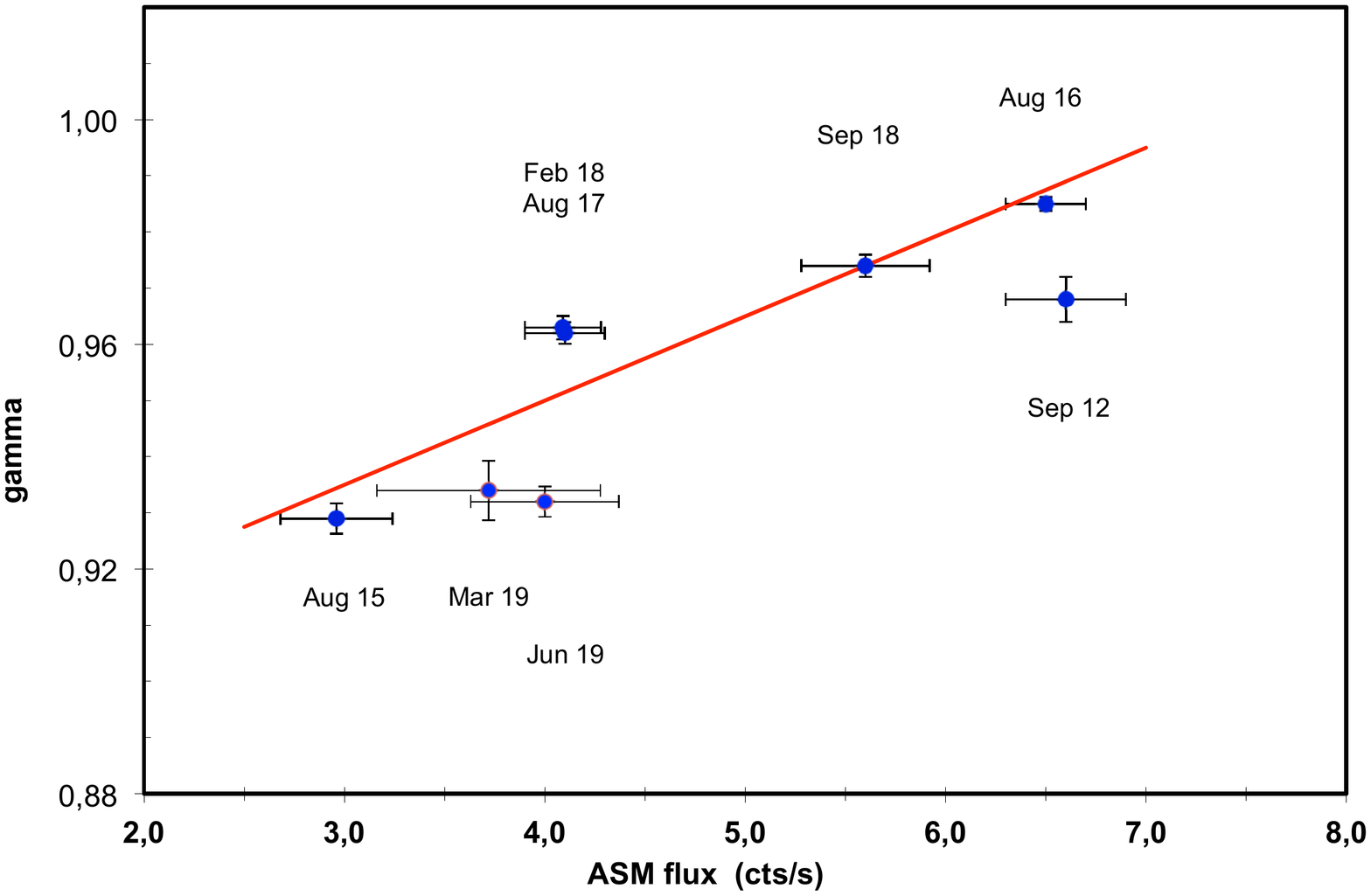}
\vspace*{-13mm}
\caption{The power law index $\Gamma$ versus flux in (ASM-cts/s). 
 The best fit line is given by the function
$\Gamma$ = $(0.965\pm0.002)$+$(0.015\pm0.001)$ $\times$ ((ASM-cts/s) - 5.0). 
This function is valid for 35-day phases up to 0.16.
The low $\Gamma$ value measured in Feb 2019 (cycle 494, see Table~\ref{tab:results}) was 
not included in this fit because the observation took place at a 35-day phase of 0.202, at which a 
lower value is expected (see text). Pearsons linear correlation coefficient is 0.53.}
\label{fig:gamma_ASM}
\end{figure}

\begin{figure}
\includegraphics[angle=90,width=10.7cm]{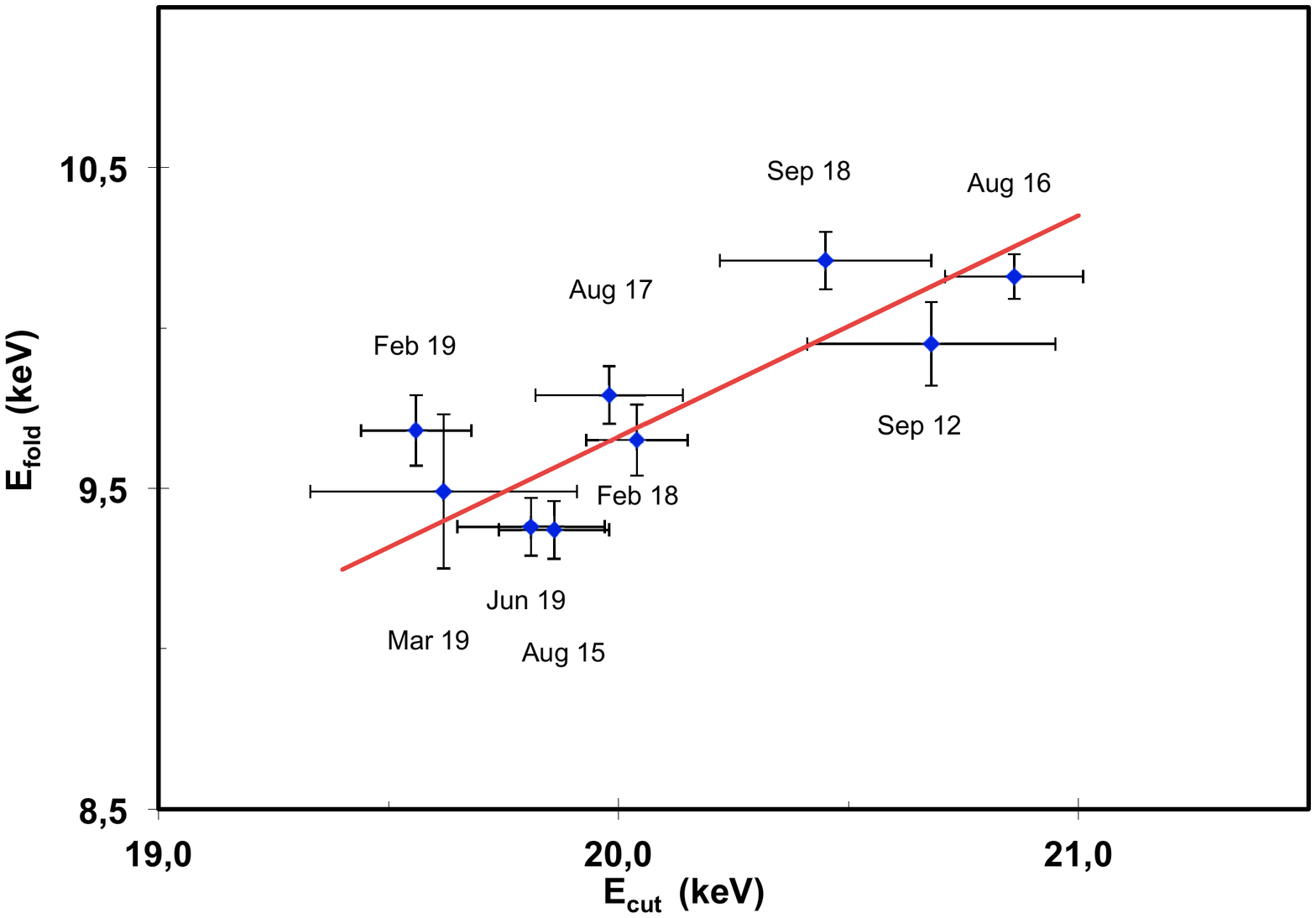}
\vspace*{-13mm}
\caption{The continuum parameter  $E_\mathrm{fold}$ versus the continuum parameter  $E_\mathrm{cut}$.
 The best fit line is given by the function
$E_\mathrm{fold}$ = $(9.66\pm0.05)+(0.69\pm0.12) \times (E_\mathrm{cut} - 20.0)$ (all values in keV).
Pearsons linear correlation coefficient is 0.82.}
\label{fig:Efold_Ecut}
\end{figure}

The systematic monitoring of Her~X-1 with \textsl{NuSTAR} over the last decade has allowed the
discovery that all continuum parameters, $E_\mathrm{cut}$,  $E_\mathrm{fold}$ and $\Gamma$ are
systematically correlated with X-ray flux. The correlations can be described by linear functions,
see Table~\ref{tab:vs_ASM}, Fig.~\ref{fig:Ecut_ASM}, Fig.~\ref{fig:Efold_ASM} and Fig.~\ref{fig:gamma_ASM}.
Since both $E_\mathrm{cut}$ and  $E_\mathrm{fold}$ depend on X-ray flux, they
correlate  with each other which is shown in Fig.~\ref{fig:Efold_Ecut}.
Normalizing $E_\mathrm{cut}$ and $E_\mathrm{fold}$ to the reference flux of 5.0~(ASM-cts/s), we find
that both normalized parameters (excluding the values of cycle 494/Feb 2019, where the 35d-phase
is very high, see Sect.~\ref{sec:35d_mod}) are consistent with a constant value over the time span 
2012-2019 (Fig.~\ref{fig:normEcut_Efold}).
In calculating the dependence of $\Gamma$ on flux we have excluded the exceptionally low value
(0.885) measured in February 2019 (cycle 494, see Table~\ref{tab:results}) because the observation
happened at a high 35-day phase of 0.202, where the flux is about 65\% of the maximum 
\textsl{Main-On} flux of this cycle. At 35-day phases beyond $\sim$0.16,  $\Gamma$ is known to
decrease \citep{Vasco_12}.  The measured increase of $\Gamma$
on flux is fairly weak but interesting because this is in disagreement with reports about a correlation 
in the opposite sense by \citet{Klochkov_etal11} (see the discussion below).

\subsection{Correlation between line and continuum parameters}

\begin{figure}
\includegraphics[angle=90,width=10.7cm]{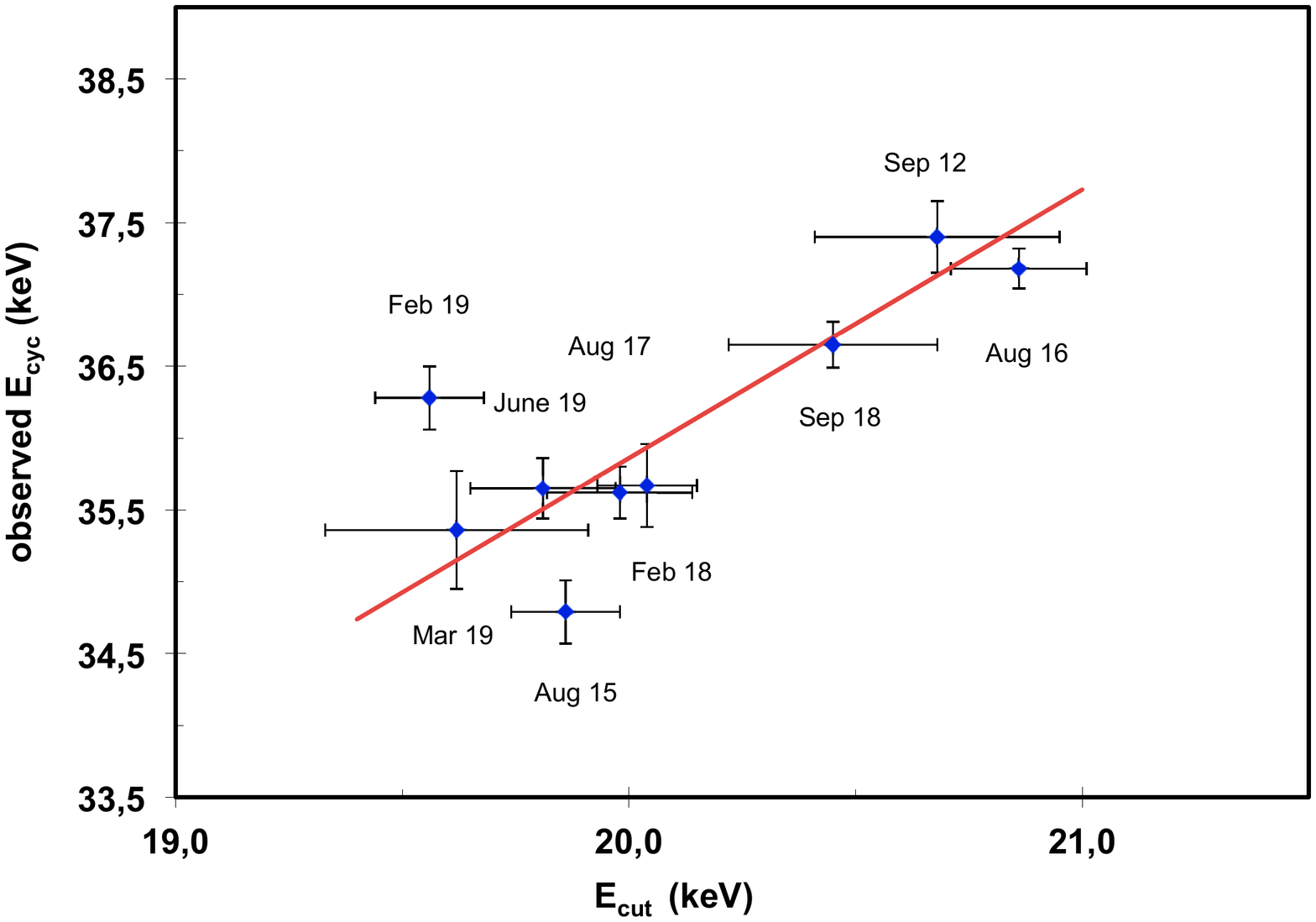}
\vspace*{-13mm}
\caption{The observed cyclotron line energy versus $E_\mathrm{cut}$.
 The best fit line is given by the function
$E_\mathrm{cyc}$ = $(35.86\pm0.11)+(1.87\pm0.36) \times (E_\mathrm{cut} - 20.0)$ (all values in keV).
Pearsons linear correlation coefficient is 0.79.}
\label{fig:Ecyc_Ecut}
\end{figure}

\begin{figure}
\includegraphics[angle=90,width=10.7cm]{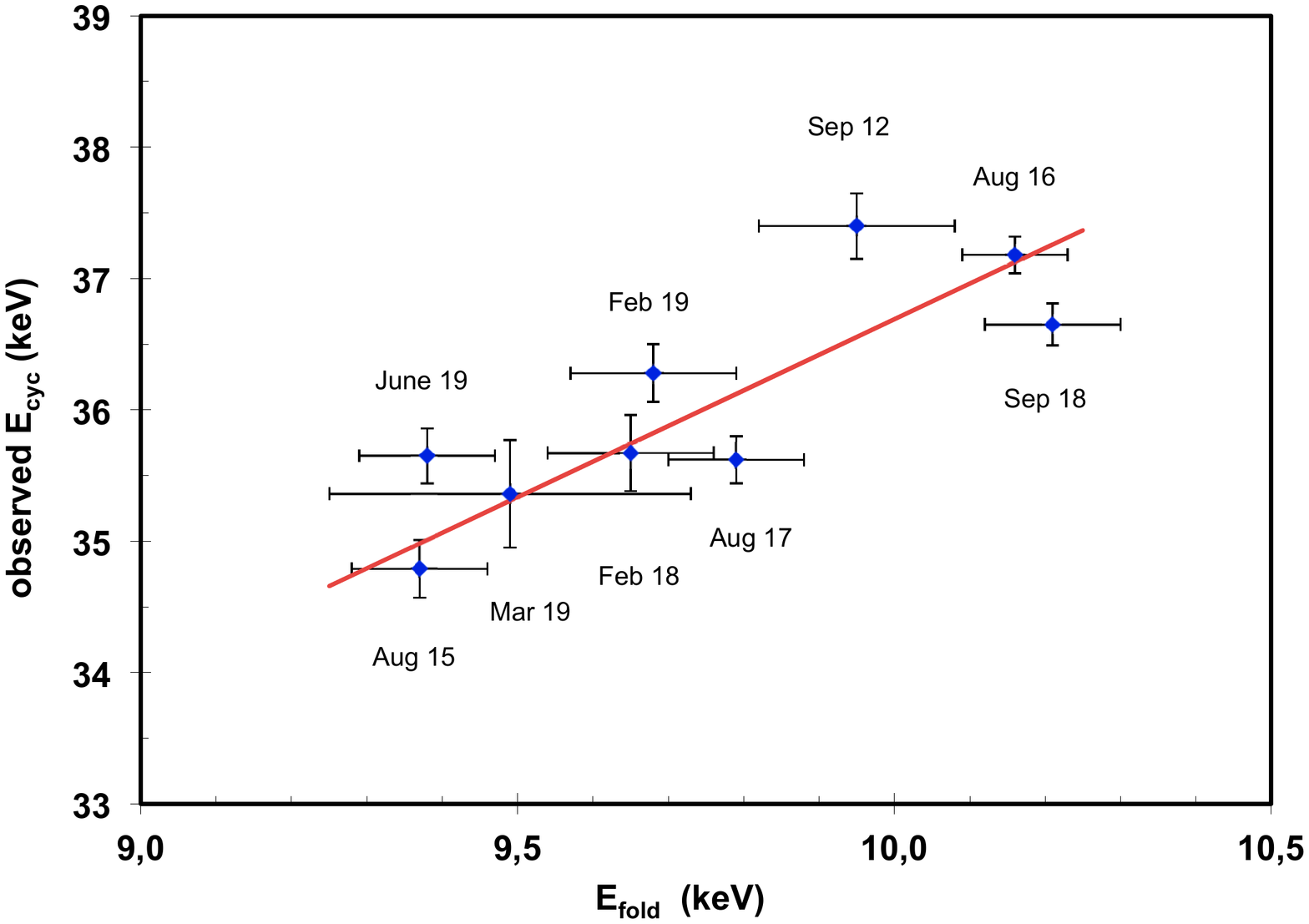}
\vspace*{-13mm}
\caption{The observed cyclotron line energy versus $E_\mathrm{fold}$.
 The best fit line is given by the function
$E_\mathrm{cyc}$ = $(36.69\pm0.13)+(2.71\pm0.39) \times (E_\mathrm{fold} - 10.0)$ (all values in keV).
Pearsons linear correlation coefficient is 0.82.}
\label{fig:Ecyc_Efold}
\end{figure}

\begin{figure}
\includegraphics[angle=90,width=10cm]{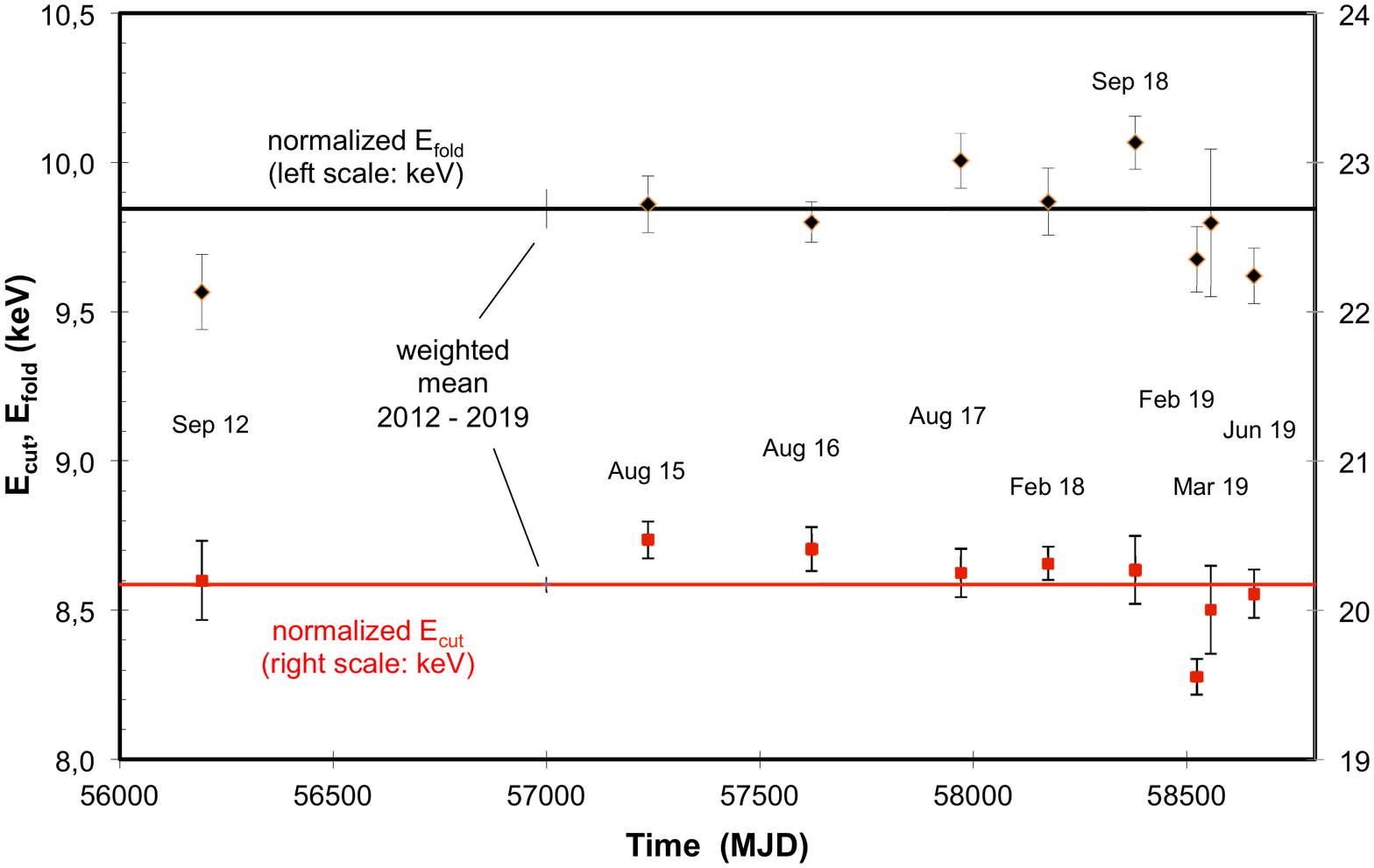}
\vspace*{-13mm}
\caption{Flux normalized values of $E_\mathrm{cut}$ (right scale) and $E_\mathrm{fold}$ (left scale)
as function of time. The normalization to a reference X-ray flux of 5~(ASM-cts/s) uses 
the linear correlations as stated in Table~\ref{tab:vs_ASM} and shown in Figs.~\ref{fig:Ecut_ASM}
and \ref{fig:Efold_ASM}. These two continuum parameters appear to be constant (see the horizontal l
ines) over the time 2012--2019, except for Feb 2019 where the 35d-phase is very high (0.202).}
\label{fig:normEcut_Efold}
\end{figure}

Since the continuum parameters $E_\mathrm{cut}$ and  $E_\mathrm{fold}$ and all cyclotron line
parameters correlate with X-ray flux, there are correlations between the line and the continuum
parameters. As two examples we list the linear correlation parameters of the observed cyclotron line 
energy and the continuum parameters in Table~\ref{tab:inter_relations} and show
them in Fig.~\ref{fig:Ecyc_Ecut} and Fig.~\ref{fig:Ecyc_Efold}.

\subsection{Dependence on phase of the 35-day modulation}
\label{sec:35d_mod}

As mentioned in the Introduction, Her~X-1 shows a regular 35-day modulation,
known since the discovery of the source by \textsl{UHURU} \citep{Tananbaum_etal72},
thought to be connected with the precession of the accretion disk providing  regular
shadowing of the X-ray source. The 35-day periodicity is also seen in the regular variations
of the shape of the pulse profiles \citep{Truemper_etal86,Staubert_etal13}, which has also 
led to the suggestion that free precession of the neutron star may play a role
\citep{Truemper_etal86,Postnov_etal13}, which is still an open question and highly debated.
In the context of this work it is of interest whether the X-ray spectra show variations with 
phase of the 35-day modulation. This is indeed the case \citep{Parmar_etal80,Mihara_etal91b,
Kuster_etal05,Vasco_12}.
\citealt{Staubert_etal14} had suggested that there is a weak modulation of the cyclotron line
energy $E_\mathrm{cyc}$ during the \textsl{Main-On} and \citealt{Vasco_12} had found a
strong variation of the power law index for 35-day phases greater than 0.16. The contribution
of the series of \textsl{NuSTAR} observations of Her~X-1, discussed here, is the following
(for the limited range provided by the data used here, see Table~\ref{tab:obs}):
1) The flux normalized $E_\mathrm{cyc}$ does not show any variation, the values are
consistent with a constant.
2) The flux normalized values of $E_\mathrm{cut}$ and $E_\mathrm{fold}$
are constant up to the second highest 35-day phase observed (0.147 for Aug 2017, cycle 478), 
then both drop to significant lower values for the highest observed 35-day phase (0.202 for
Feb 2019, cycle 494). 
3) The normalized power law index $\Gamma$ behaves as the other 
two continuum parameters and drops to the lowest value (0.885) for the Feb 2019 observation
at 35-day phase 0.202. 

\section{Discussion}

Correlations between spectral parameters
were first discovered by comparing spectral parameters determined for different objects
\citep{MakishimaMihara_92,Makishima_etal99,Coburn_etal02}. We now see, most prominently
(but not only) in Her~X-1, that strong correlations between spectral parameters, both continuum
and line parameters, exist for individual objects, particularly related to changes in X-ray luminosity.

\subsection{Correlation between line parameters}

The linear correlation between the width of the cyclotron line and its centroid energy 
(Fig.~\ref{fig:sigcyc_Ecyc}) is expected from the fact that the line suffers a Doppler-broadening due to 
the thermal motion of the electron gas at a temperature $kT_e$. Applying the general formula for a 
Doppler-broadened line (of central energy E) to the resonant cyclotron scattering of photons on electrons, 
we write $\sigma$~=~E~$(kT_e/m_e~c^{2})^{1/2}$.\footnote{The general formula for a Doppler-broadened 
line is FWHM = E~$\times$~$(8~ln2~kT_e~/~m_e~c^{2})^{1/2}$, with E being the line energy and
FWHM = $\sigma$~$(8~ln2)^{1/2}$~=~2.356~$\sigma$ being the full width-at-half-maximum 
(see, e.g., K.R. Lang, Astrophysical Formulae, Springer).}
Because electrons in a strong magnetic field can move freely only in one dimension (along the field lines) we need to
multiply by $cos~\theta$, with $\theta$ being the the angle between the viewing direction and the magnetic field lines.
The temperature of the electron gas is $kT_e$ and the rest mass of the electron is $m_e~c^{2}$$\approx$~511\,keV,
so the electron temperature can be estimated to 
$kT_e$~[keV]~$\approx~$511~($\sigma / E_\mathrm{cyc})^{2} / |cos^{2}\Theta|$.
The line broadening effect was already pointed out by \cite{Truemper_etal77} when the 
discovery of the first cyclotron line  - in Her X-1 - was reported
(see also \citealt{MeszarosNagel_85,Orlandini_etal98}). It was then observationally confirmed
when different cyclotron line energies and associated widths in several X-ray binaries were measured
\citep{Makishima_etal99,Coburn_etal02}. Recently, it became possible to observe such correlations in
individual sources, when the CRSF energy as well as the line width change with varying flux. Here we report
the measurement for Her~X-1: $\sigma$ = 5.26~+~0.60~($E_\mathrm{cyc}$~-~36.0) (all in keV).
For small $\theta$ (cos~$\theta$~$\sim$~1), the calculated $kT_e$ ranges from 8.3\,keV to 13.5\,keV,
for flux values from 3 to 7 ASM-cts/s, respectively. This is very close to 10\,keV which is the typical value 
of the continuum parameter $E_\mathrm{fold}$, often taken as the temperature of the plasma emitting the continuum.
It is tempting to conclude that in Her X-1 we most likely see a pencil beam rather than a fan beam.

The fact that the relative line width $\sigma / E_\mathrm{cyc}$ increases with increasing flux 
(Fig.~\ref{fig:relat_sig_ASM}) means that the electron temperature $kT_e$ increases with increasing flux.
And so does $E_\mathrm{fold}$, as expected for an increasing accretion rate.
However, the magnitude of the increase in $kT_e$ is $\sim$12\,\% per unit flux, significantly stronger 
than the increase in $E_\mathrm{fold}$ with only $\sim$2\,\%. If $E_\mathrm{fold}$ is indeed a measure
of the continuum temperature, then the electron temperature is increasing significantly faster.

\begin{figure*}
\includegraphics[angle=90,width=18cm]{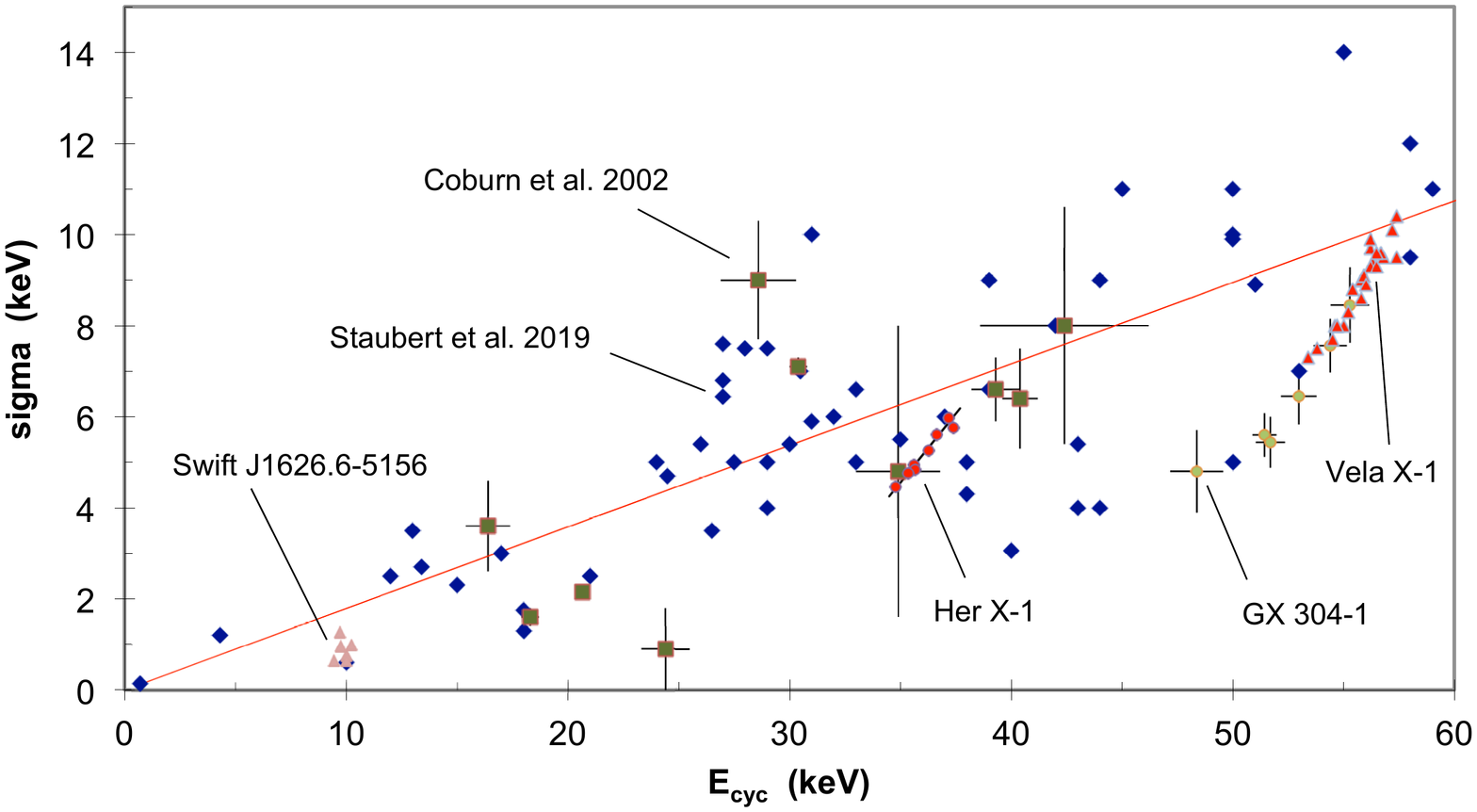}
\vspace*{-1.5cm}
\caption{Correlation between the width (sigma) and the centroid energy of cyclotron
lines for different X-ray binaries. The blue rhombs are data from the compilation in
\citet{Staubert_etal19}, the filled squares (with uncertainties) are from \citet{Coburn_etal02}.
The red line is a fit to these two data sets defining $\sigma$~=~0.18~$E_\mathrm{cyc}$.
The individual sources shown are: GX~304$-$1 \citep{Klochkov_etal15},
Swift~J1626.6$-$5156 \citep{DeCesar_etal13}, Vela~X-1 \citep{LaParola_etal16} and Her~X-1 
(same as Fig.~\ref{fig:sigcyc_Ecyc}).}
\label{fig:sig_vs_Ecyc_all}
\end{figure*}

The general dependence of $\sigma$ on $E_\mathrm{cyc}$ for the known cyclotron line objects 
is demonstrated in Fig.~\ref{fig:sig_vs_Ecyc_all} (up to $E_\mathrm{cyc}$~=~60\,keV), where 
pairs of $\sigma$ and $E_\mathrm{cyc}$ values are shown, taken from \citet{Coburn_etal02} and 
 \citet{Staubert_etal19} (their Table~A.5), together with a few individual objects. The central line
through the rather scattered data defines a slope of $\sim$0.18\,keV/keV, corresponding to a mean
$kT_e$ of $\sim$16\,keV for $\Theta$~=~0. For a few objects we have now observed variations of 
both the line energy and the line width (physically introduced by a changing X-ray flux): Her~X-1 
(\citealt{Staubert_etal07} and this work), GX~304$-$1 
\citep{Klochkov_etal15,Malacaria_etal15,Rothschild_etal17}, Vela~X-1 \citep{LaParola_etal16} 
(the first harmonic) and Swift~J1626.6$-$5156 \citep{DeCesar_etal13} (see Fig.~\ref{fig:sig_vs_Ecyc_all}).
For these few objects the absolute values of the relative widths are all rather small (they are all below 
the red line in Fig.~\ref{fig:sig_vs_Ecyc_all}, but the variation d($\sigma$)~/~d($E_\mathrm{cyc}$), 
is significantly steeper (e.g., for Her~X-1: 0.60\,keV/keV), than for the complete ensemble.

We would like to stress that the X-ray flux (physically the mass accretion rate) is a fundamental 
parameter that seems to influence all spectral parameters. 
Apart from the line position and width, also its strength, 
its depth and its relative width are positively correlated with flux (see Table~\ref{tab:vs_ASM}). 
The same is true for the continuum parameters $E_\mathrm{cut}$ and $E_\mathrm{fold}$ 
(Fig.~\ref{fig:Ecut_ASM} and Fig.~\ref{fig:Efold_ASM}), and - surprisingly (even if weak) -
for the power law index $\Gamma$. We find that $\Gamma$ increases with flux (by 1.8\%
for an increase in flux by a factor of two), while \citet{Klochkov_etal11}, in pulse-amplitude-resolved 
spectroscopy of Her~X-1, found the opposite trend (by 5.6\%) - always in combination with an increase 
of $E_\mathrm{cyc}$ with flux.
A solution may be given by \citet{Postnov_etal15} who showed (for several sources, but unfortunately 
not Her~X-1) that the spectral hardness correlates with X-ray flux, consistent with \citet{Klochkov_etal11}, 
but only up to a (source dependent) luminosity around a few times $10^{37}$\,erg/s, after which 
the correlation stops, or even reverses. A luminosity of a few times $10^{37}$\,erg/s is considered 
to be close to the border between the sub- and super-critical accretion regimes \citep{Becker_etal12}
at which the trend for these correlations reverses. Her~X-1 is probably operating close to the
critical luminosity, and the turning point for a reversal may actually be slightly different between
the respective correlations. 

An interesting correlation, first found by \citet{Coburn_etal02} among a group of X-ray binaries,
namely the relative line width $\sigma_\mathrm{cyc}$ / $E_\mathrm{cyc}$ as function of the
optical depth $\tau$, is also realized in individual objects like 4U~1538$-$52 \citep{Rodes-Roca_etal08} and
 in Her~X-1 (Fig.~\ref{fig:relat_sig_tau}). \citet{Coburn_etal02} and \citet{Rodes-Roca_etal08}
note that simple theoretical models of cyclotron line generation actually predict the opposite dependence
\citep{Isenberg_etal98,ArayaHarding_99}.

Regarding inter-correlations between spectral parameters, it has been a general
worry about how large the influence is of purely mathematical correlations introduced
in the fitting process. \citet{Coburn_etal02} have tried to answer this question by performing
Monte Carlo simulations, carefully designed for different types of correlations. Their conclusion 
was that  formal correlations were not significant and that it was safe to conclude that
the observed correlations are a true physical effect. We have used the  Markov Chain Monte 
Carlo (MCMC) procedure offered in XSPEC to investigate the same question and present a 
corresponding analysis in Appendix~A. We are confident that the observed correlations are 
indeed physical.

\subsection{Physics behind changes of $E_\mathrm{cyc}$}

The centroid energy of the cyclotron line in Her~X-1 has been observed to change 
with respect to the following parameters \citep{Staubert_etal14}: \\
- pulse phase: 20\% ((max-min)/mean) \citep{Voges_etal82,Vasco_etal13,Staubert_etal14};\\
- X-ray flux: 6.5\% for a change of flux by a factor of two (here and \citealt{Staubert_etal07}); \\
- elapsed time: constant around 35\,keV from the discovery in 1975 to 1990, 
jump upward 1991--1994 from $\sim$35\,keV to beyond 40\,keV ($\sim$20\%)
\citep{Gruber_etal01,Staubert_etal07}, followed by a well measured decay until  
$\sim$2012 down to $\sim$37 keV (10\% over 16 years)  
(here and \citealt{Staubert_etal14,Staubert_etal16}); \\
- and possibly a change by 1\,keV or less with phase of the 35-day on-off-cycle
\citep{Staubert_etal14}, which we do not confirm here.

The variation with pulse phase is believed to be due to the changing viewing angle under 
which the emission regions are seen \citep{Schoenherr_etal07}. 
\citealt{Vasco_12}\footnote{PhD thesis by Davide Vasco, 2012, University of Tübingen, 
Germany, http://nbn-resolving.de/urn:nbn:de:bsz:21-opus-63466} 
has shown that the above discussed 
correlation between the width and the centroid energy of the cyclotron line (and the dependence
of both on flux) is also valid when certain pulse phases are analyzed (e.g., the line energy and the width
are both at a maximum around the peak of the main pulse).
We will not further discuss this phenomenon here, but concentrate on the dependence  of
the pulse phase averaged cyclotron line energy $E_\mathrm{cyc}$ on X-ray flux and on time.

\subsubsection{Changes of $E_\mathrm{cyc}$ with flux}

With respect to the physics at work behind the positive correlation of the pulse phase
averaged cyclotron line energy with flux, we refer to discussions presented earlier by
\citet{Staubert_etal07,Staubert_etal14,Staubert_etal16,Staubert_etal17,Ji_etal19}, as 
well as the theoretical work by \cite{Becker_etal12}
(see also the summary and references given in the review by \citet{Staubert_etal19}).
Here detailed modeling of the physics is necessary: what is the mechanism of 
deceleration of the accreted material? Is it due to simple Coulomb scattering or the
generation of radiative shocks? What kind of accretion rate is necessary to generate
such shocks and at which height above the neutron star surface would they form?
What is the configuration of the magnetic field, most likely influenced by the in-falling
material?
It has become popular to talk about `accretion regimes'' \citep[e.g.,][]{Becker_etal12}:
 `sub-critical" and  `super-critical" accretion, most likely separated by a critical
 luminosity of the order of a few times $10^{37}$\,erg/s. Generally, the  `sub-critical"
 accretion is associated with a positive $E_\mathrm{cyc}$/$L_{x}$ correlation,
 and the `super-critical" accretion with a negative correlation
 \citep{Becker_etal12,Mushtukov_etal15a}. Recently, a model involving a collisionless
shock was developed that also explains the deviation from a pure
linear dependence (a ``roll-off''), as observed in GX~304$-$1
\citep{Rothschild_etal17,Vybornov_etal17}. A more detailed discussion is presented in
\citet{Staubert_etal17}.

Alternatively, or in combination with a height-related effect, the observed variations could be due 
to changes of the configuration of the magnetic field when the accretion rate varies. 
As \citet{Mukherjee_etal13b,Mukherjee_etal14}
have shown, the usually assumed dipole structure of the magnetic field is significantly altered
when the mass accretion rate changes. Close to the magnetic poles, a higher accretion rate can lead to 
a significant increase of the density of field lines at the outer circumference of the accretion
mound when the in-falling material pushes matter and field lines from the center radially
outward. See also the Discussion in \citet{Bala_etal20}.

\subsubsection{Changes of $E_\mathrm{cyc}$ with time}

The dependence of $E_\mathrm{cyc}$ on time is even less well understood. 
With regard to the long-term decay of $E_\mathrm{cyc}$, we think that it is either a geometric 
displacement of the emission region or a change in the local field configuration e.g., as calculated 
by \citet{MukhBhatt_12}, rather than a change 
in the strength of the underlying global dipole. \citet{Staubert_14} has suggested that the observed 
change of E$_\mathrm{cyc}$ may be connected to a slight imbalance between ``gain" and ``loss"
of accreted material, such that the structure of the accretion mound changes with time. With an accretion 
rate of ${\sim}10^{17}$\,g/s a variation on relatively short time scales does not seem implausible.
If the ``gain" is slightly larger than the ``loss", material would slowly accumulate in the mound,
possibly increasing its height or changing the local field structure, which might be the reason for the 
long-term reduction in $E_\mathrm{cyc}$. One might expect, that this reduction can find a
natural end (e.g., when the excess mass and the associated pressure in the accretion 
mound becomes too large), such that a forced outflow of material to larger areas of the NS surface
causes a re-adjustment of the accretion mound back to the un-perturbed configuration. This could
be a relatively fast and catastrophic event -- possibly explaining the rather sudden upward jump in 
$E_\mathrm{cyc}$ observed between 1990 and 1993. 
The time period 1990-2012 sets an apparent time scale of instability: a few years of very fast
change -- the increase in $E_\mathrm{cyc}$ by $\sim$5\,keV --, followed by $\sim$16\,years of decay
down to the original level. A corresponding time scale for stability is not yet known.
Future observations should search for indications for an increase or even a new upward jump in 
$E_\mathrm{cyc}$. Since an upward movement could be rather fast (similar to the earlier event), 
it is important to observe as regularly as possible, in order not to miss such an event again.

\section{Summary}

Her~X-1 has been well monitored during the last decade, mostly by \textsl{NuSTAR}, but also
by \textsl{INTEGRAL} and \textsl{Swift}/BAT,  and more recently by \textsl{Insight}-HXMT  
\citep{Xiao_etal19} and \textsl{Astrosat} \citep{Bala_etal20}.
Her~X-1 is the only highly magnetized accreting pulsar for which repeated observations 
over longer periods of time exist. This has provided the base for the discovery of new phenomena, 
like the dependence of the cyclotron line energy (as well as almost all spectral parameters) on flux 
and the long-term decay of the cyclotron line energy over nearly 20 years. 
Both of these phenomena have meanwhile been seen in other accreting X-ray binary pulsars 
(see the review by \citealt{Staubert_etal19}). 

The results of nine \textsl{NuSTAR} observations of Her~X-1 between 2012 and 2019 are the
following: \\
- The dependence of the cyclotron line energy on X-ray flux, discovered in 2007, is confirmed
and measured with high precision. \\
- The flux-normalized cyclotron line energy is constant since (at least) $\sim$2012. The previously 
reported long-term decay has ended. \\
- All cyclotron line parameters - the line energy $E_\mathrm{cyc}$, the width $\sigma$, the
strength, the optical depth $\tau$ and the relative width - show a positive and linear correlation to X-ray flux. \\
- The former statement also means that all cyclotron line parameters correlate positively
and linearly with one-another. \\
- The continuum parameters $E_\mathrm{cut}$ and $E_\mathrm{fold}$ correlate positively and linearly 
with X-ray flux. The flux-normalized continuum parameters are consistent with constant values since 2012.
The third continuum parameter, the power law index $\Gamma$, shows a weak positive
correlation with flux. This is the opposite of what was seen before in a pulse-amplitude-resolved analysis,
which may have a somewhat different meaning. The interpretation of the different observed correlations with 
regard to the prevailing accretion regimes (sub- or supercritical) is not so simple. \\
- We have learned that there are correlations between continuum parameters and line parameters. \\
- The correlations of the line and continuum parameters with X-ray flux and among each other are considered 
to reflect true physical correlations, which have yet to be investigated and explained by theoretical modeling. 
As shown in the Appendix, the analysis of the purely mathematical correlation between fitting parameters has 
led to the conclusion that those are not significant.

We urge that the source continues to be observed regularly.  For 2020 this seems to be secured 
through already planned observations (partly simultaneous) between \textsl{INTEGRAL}, \textsl{NuSTAR},
\textsl{XMM-Newton} and \textsl{Insight}-HXMT.
At the same time, it would be very important that theoretical models be developed further.

\begin{acknowledgements}
We would like to acknowledge the dedication of all the people who have contributed to the 
great success of all relevant missions, in particular \textsl{NuSTAR}.
We especially thank the ``schedulers'', foremost  Karl Forster, for his efforts with respect to the 
non-standard scheduling of the observations of Her~X-1. We acknowledge the historical 
contributions to the subject at hand by Dmitry Klochkov. 
We thank the anonymous referee for valuable comments.
\end{acknowledgements}

\appendix

\section{Appendix}

In the main text, we reported the correlations between spectral parameters observed from nine \textit{NuSTAR} 
observations in 2012-2019. Here we ask how much model degeneracies during the spectral fitting contribute to
theses correlations.
We investigate this  through Monte Carlo simulations. In practice, we adopted the best-fitting parameters of the 
\textsl{NuSTAR} observation in September 2018 as a reference model (with a flux of 5.6\,(ASM-cts/s) this observation 
is close to the center of the flux range encountered between 2012 and 2019). With the statistics and the spectral 
parameters from this observation as an input model, we performed two different simulations (with $10^4$ events each): 
first, a Markov Chain Monte Carlo (MCMC) simulation (provided by XSPEC\footnote{https://heasarc.gsfc.nasa.gov/xanadu/xspec/}) 
which makes use of the Goodman-Weare algorithm\footnote{https:///ui.adsabs.harvard.edu/abs/2010CAMCS...5...65G/abstract} 
(see also \citet{Foreman-Mackey_13}), and second producing simulated spectra using the \textit{fakeit} 
command in {\sc XSPEC} with subsequently fitting these spectra. The two methods provided consistent results,
we show the MCMC simulation in Fig.~\ref{fig:MCMC}.

\begin{figure}
        \centering
        \includegraphics[angle=90,width=10cm]{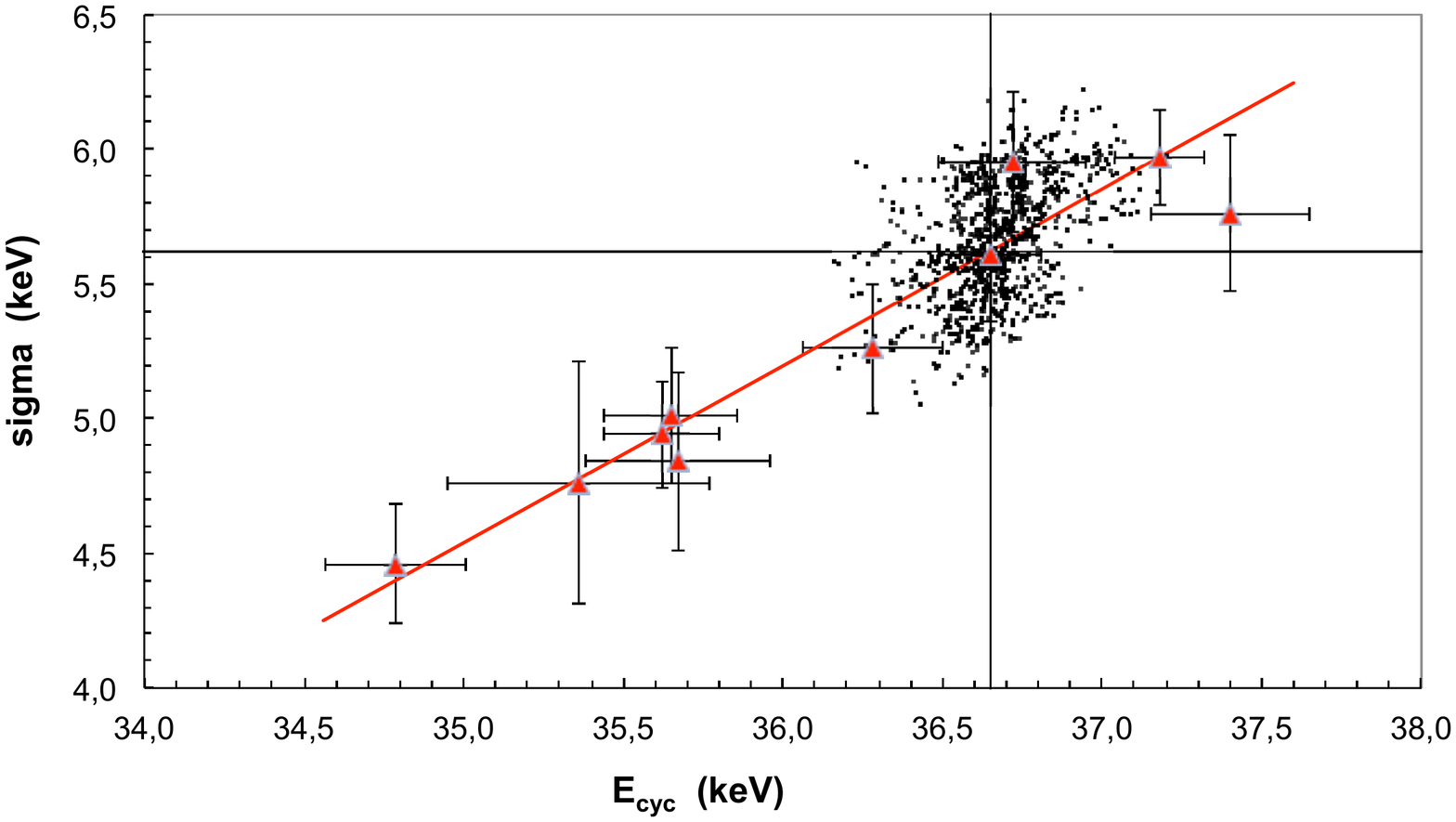}
\vspace{-15mm}
        \caption{The first 1000 sigma / $E_{\rm cyc}$  pairs of the MCMC simulation together with the
        measured correlation as shown in Fig.~\ref{fig:sigcyc_Ecyc}.}
                \label{fig:MC_sig_Ecyc}
\end{figure}

\begin{figure}
        \centering
        \includegraphics[angle=90,width=10cm]{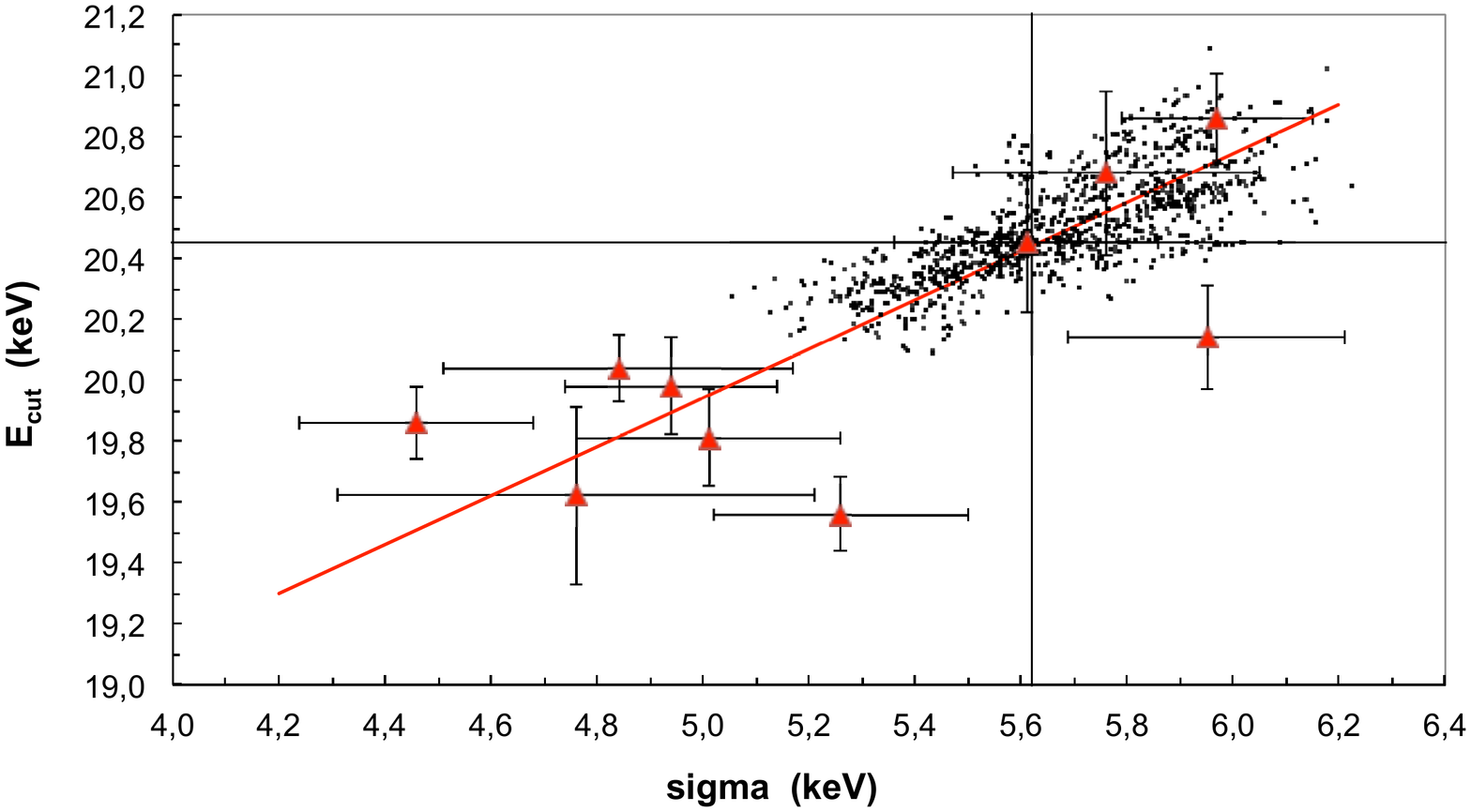}
\vspace{-15mm}
        \caption{The first 1000 $E_{\rm cut}$ / sigma  pairs of the MCMC simulation together with the
        measured correlation.}
                \label{fig:MC_Ecut_sig}
\end{figure}

As expected, most of the correlations between the parameters are weak, as is evident from the circular shape
of the two-dimensional distributions. As an example we show the scatter plot of $\rm \sigma_{\rm cyc}$ versus
$E_{\rm cyc}$ of the first 1000 simulated MCMC spectra in Fig.~\ref{fig:MC_sig_Ecyc}. The range of observed
$E_{\rm cyc}$ is a factor $\sim$10 larger than the corresponding full width at half maximum (FWHM) of the
simulated $E_{\rm cyc}$ distribution (for $\rm \sigma_{\rm cyc}$ the factor is $\sim$3.3). This shows that any
model degeneracy has only a minor contribution to the overall correlation.
There are three distributions that are elongated (under roughly 45 degrees) indicating stronger correlations: 
$D_{\rm cyc}$\footnote{$D_{\rm cyc}$ is called \textsl{strength} in the main text} versus $\rm \sigma_{\rm cyc}$,  
$E_{\rm cut}$ versus $\rm \sigma_{\rm cyc}$ and $E_{\rm cut}$ versus $D_{\rm cyc}$. Even here the corresponding
factors (FWHM / observed range) are between two and four. The correlation between $\rm \sigma_{\rm cyc}$
and $D_{\rm cyc}$ is actually given through the definition: $D_{\rm cyc}$ = $\sigma \times \tau \times \sqrt{2\pi}$.
In Fig.~\ref{fig:MC_Ecut_sig} we show the degeneracy between the continuum parameter $E_{\rm cut}$ and
the line parameter $\rm \sigma_{\rm cyc}$. We conclude that the physical correlations discussed are real.

\begin{figure*}
        \centering
        \includegraphics[width=18cm]{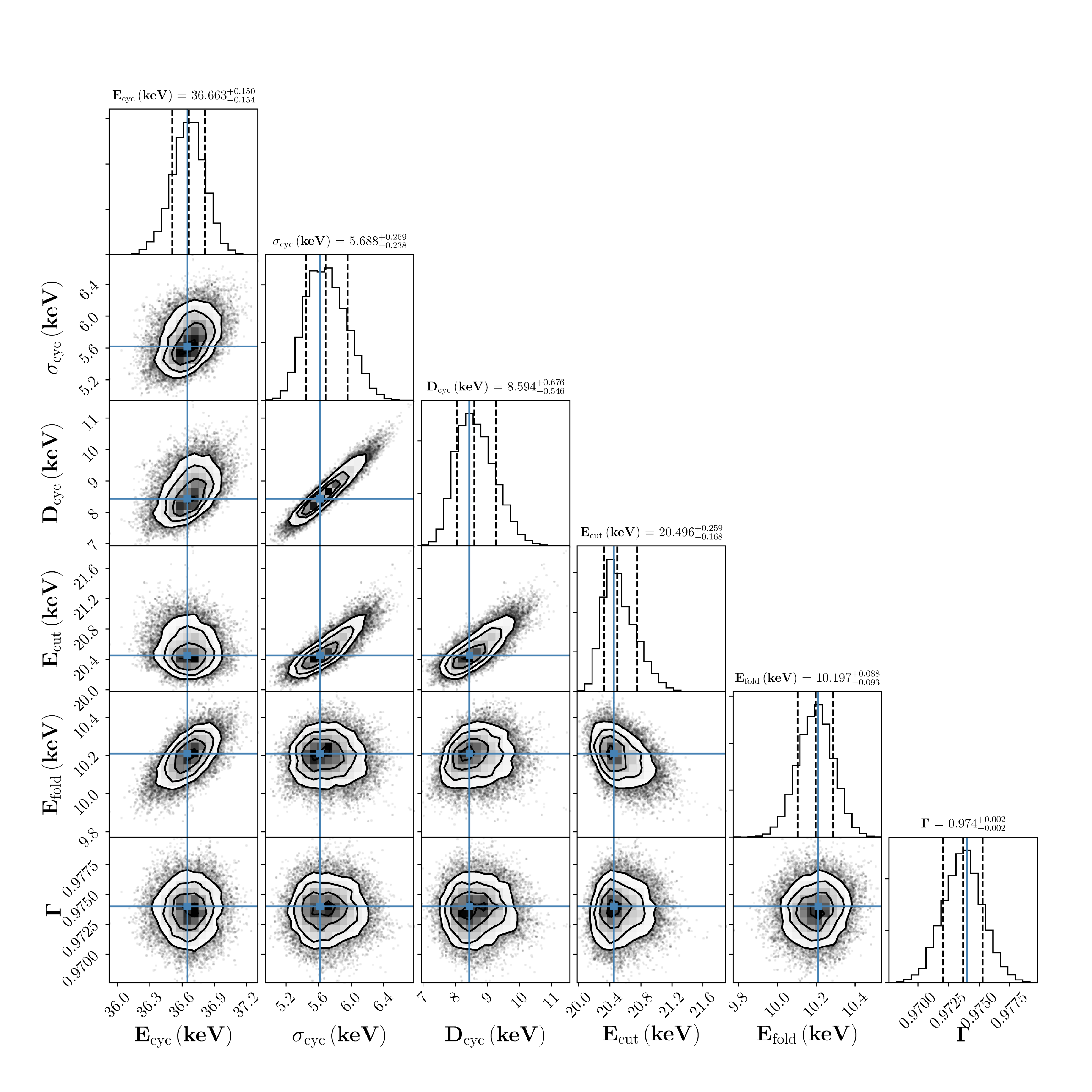}
        \caption{The contours represent two-dimensional distributions of parameters obtained from Monte Carlo simulations at the 
        significance level of 1, 2 and 3\,$\sigma$, and the histograms are distributions for each variable. The blue lines 
        are the input parameters used during the simulations.}
        \label{fig:MCMC}
\end{figure*}


\begin{thebibliography}{63}
\expandafter\ifx\csname natexlab\endcsname\relax\def\natexlab#1{#1}\fi

\bibitem[{Araya \& Harding(1999)}]{ArayaHarding_99}
Araya, R. \& Harding, A. 1999, ApJ, 517, 334

\bibitem[{{Bala} {et~al.}(2020){Bala}, {Bhattacharya}, {Staubert}, \&
  {Maitra}}]{Bala_etal20}
{Bala}, S., {Bhattacharya}, D., {Staubert}, R., \& {Maitra}, C. 2020, \mnras,
  497, 1029

\bibitem[{{Becker} {et~al.}(2012){Becker}, {Klochkov}, {Sch{\"o}nherr},
  {Nishimura}, {Ferrigno}, {Caballero}, {Kretschmar}, {Wolff}, {Wilms}, \&
  {Staubert}}]{Becker_etal12}
{Becker}, P.~A., {Klochkov}, D., {Sch{\"o}nherr}, G., {et~al.} 2012, \aap, 544,
  A123

\bibitem[{{Caballero} \& {Wilms}(2012)}]{CaballeroWilms_12}
{Caballero}, I. \& {Wilms}, J. 2012, \memsai, 83, 230

\bibitem[{{Coburn} {et~al.}(2002){Coburn}, {Heindl}, {Rothschild}, {Gruber},
  {Kreykenbohm}, {Wilms}, {Kretschmar}, \& {Staubert}}]{Coburn_etal02}
{Coburn}, W., {Heindl}, W.~A., {Rothschild}, R.~E., {et~al.} 2002, ApJ, 580,
  394

\bibitem[{{DeCesar} {et~al.}(2013){DeCesar}, {Boyd}, {Pottschmidt}, {Wilms},
  {Suchy}, \& {Miller}}]{DeCesar_etal13}
{DeCesar}, M.~E., {Boyd}, P.~T., {Pottschmidt}, K., {et~al.} 2013, \apj, 762,
  61

\bibitem[{{Foreman-Mackey} {et~al.}(2013){Foreman-Mackey}, {Hogg}, {Lang}, \&
  {Goodman}}]{Foreman-Mackey_13}
{Foreman-Mackey}, D., {Hogg}, D.~W., {Lang}, D., \& {Goodman}, J. 2013, \pasp,
  125, 306

\bibitem[{{F{\"u}rst} {et~al.}(2013){F{\"u}rst}, {Grefenstette}, {Staubert},
  {Tomsick}, {Bachetti}, {Barret}, {Bellm}, {Boggs}, {Chenevez}, {Christensen},
  {Craig}, {Hailey}, {Harrison}, {Klochkov}, {Madsen}, {Pottschmidt}, {Stern},
  {Walton}, {Wilms}, \& {Zhang}}]{Fuerst_etal13}
{F{\"u}rst}, F., {Grefenstette}, B.~W., {Staubert}, R., {et~al.} 2013, \apj,
  779, 69

\bibitem[{{Gerend} \& {Boynton}(1976)}]{GerendBoynton_76}
{Gerend}, D. \& {Boynton}, P.~E. 1976, ApJ, 209, 562

\bibitem[{{Gruber} {et~al.}(2001){Gruber}, {Heindl}, {Rothschild}, {Coburn},
  {Staubert}, {Kreykenbohm}, \& {Wilms}}]{Gruber_etal01}
{Gruber}, D.~E., {Heindl}, W.~A., {Rothschild}, R.~E., {et~al.} 2001, ApJ, 562,
  499

\bibitem[{{Harrison} {et~al.}(2013){Harrison}, {Craig}, {Christensen},
  {Hailey}, {Zhang}, {Boggs}, {Stern}, {Cook}, {Forster}, {Giommi},
  {Grefenstette}, {Kim}, {Kitaguchi}, {Koglin}, {Madsen}, {Mao}, {Miyasaka},
  {Mori}, {Perri}, {Pivovaroff}, {Puccetti}, {Rana}, {Westergaard}, {Willis},
  {Zoglauer}, {An}, {Bachetti}, {Barri{\`e}re}, {Bellm}, {Bhalerao},
  {Brejnholt}, {Fuerst}, {Liebe}, {Markwardt}, {Nynka}, {Vogel}, {Walton},
  {Wik}, {Alexander}, {Cominsky}, {Hornschemeier}, {Hornstrup}, {Kaspi},
  {Madejski}, {Matt}, {Molendi}, {Smith}, {Tomsick}, {Ajello}, {Ballantyne},
  {Balokovi{\'c}}, {Barret}, {Bauer}, {Blandford}, {Brandt}, {Brenneman},
  {Chiang}, {Chakrabarty}, {Chenevez}, {Comastri}, {Dufour}, {Elvis}, {Fabian},
  {Farrah}, {Fryer}, {Gotthelf}, {Grindlay}, {Helfand}, {Krivonos}, {Meier},
  {Miller}, {Natalucci}, {Ogle}, {Ofek}, {Ptak}, {Reynolds}, {Rigby},
  {Tagliaferri}, {Thorsett}, {Treister}, \& {Urry}}]{Harrison_etal13}
{Harrison}, F.~A., {Craig}, W.~W., {Christensen}, F.~E., {et~al.} 2013, \apj,
  770, 103

\bibitem[{{Heindl} {et~al.}(2004){Heindl}, {Rothschild}, {Coburn}, {Staubert},
  {Wilms}, {Kreykenbohm}, \& {Kretschmar}}]{Heindl_etal04}
{Heindl}, W.~A., {Rothschild}, R.~E., {Coburn}, W., {et~al.} 2004, in American
  Institute of Physics Conference Series, Vol. 714, X-ray Timing 2003: Rossi
  and Beyond, ed. P.~{Kaaret}, F.~K. {Lamb}, \& J.~H. {Swank}, 323--330

\bibitem[{{Isenberg} {et~al.}(1998){Isenberg}, {Lamb}, \&
  {Wang}}]{Isenberg_etal98}
{Isenberg}, M., {Lamb}, D.~Q., \& {Wang}, J.~C.~L. 1998, \apj, 493, 154

\bibitem[{{Ji} {et~al.}(2019){Ji}, {Staubert}, {Ducci}, {Santangelo}, {Zhang},
  \& {Chang}}]{Ji_etal19}
{Ji}, L., {Staubert}, R., {Ducci}, L., {et~al.} 2019, \mnras, 484, 3797

\bibitem[{{Klochkov} {et~al.}(2008){Klochkov}, {Staubert}, {Postnov},
  {Shakura}, {Santangelo}, {Tsygankov}, {Lutovinov}, {Kreykenbohm}, \&
  {Wilms}}]{Klochkov_etal08}
{Klochkov}, D., {Staubert}, R., {Postnov}, K., {et~al.} 2008, A\&A, 482, 907

\bibitem[{{Klochkov} {et~al.}(2015){Klochkov}, {Staubert}, {Postnov}, {Wilms},
  {Rothschild}, \& {Santangelo}}]{Klochkov_etal15}
{Klochkov}, D., {Staubert}, R., {Postnov}, K., {et~al.} 2015, \aap, 578, A88

\bibitem[{{Klochkov} {et~al.}(2011){Klochkov}, {Staubert}, {Santangelo},
  {Rothschild}, \& {Ferrigno}}]{Klochkov_etal11}
{Klochkov}, D., {Staubert}, R., {Santangelo}, A., {Rothschild}, R.~E., \&
  {Ferrigno}, C. 2011, \aap, 532, A126

\bibitem[{{Klochkov} {et~al.}(2006){Klochkov}, {Shakura}, {Postnov},
  {Staubert}, {Wilms}, \& {Ketsaris}}]{Klochkov_etal06}
{Klochkov}, D.~K., {Shakura}, N.~I., {Postnov}, K.~A., {et~al.} 2006, Astronomy
  Letters, 32, 804

\bibitem[{{Kuster} {et~al.}(2005){Kuster}, {Wilms}, {Staubert}, {Heindl},
  {Rothschild}, {Shakura}, \& {Postnov}}]{Kuster_etal05}
{Kuster}, M., {Wilms}, J., {Staubert}, R., {et~al.} 2005, A\&A, 443, 753

\bibitem[{{La Parola} {et~al.}(2016){La Parola}, {Cusumano}, {Segreto}, \&
  {D'A{\`i}}}]{LaParola_etal16}
{La Parola}, V., {Cusumano}, G., {Segreto}, A., \& {D'A{\`i}}, A. 2016, \mnras,
  463, 185

\bibitem[{{Makishima} \& {Mihara}(1992)}]{MakishimaMihara_92}
{Makishima}, K. \& {Mihara}. 1992, {Frontiers O X-ray Astronomy} (Universal
  Academy Press Inc, Tokyo), 23

\bibitem[{{Makishima} {et~al.}(1999){Makishima}, {Mihara}, {Nagase}, \&
  {Tanaka}}]{Makishima_etal99}
{Makishima}, K., {Mihara}, T., {Nagase}, F., \& {Tanaka}, Y. 1999, \apj, 525,
  978

\bibitem[{{Malacaria} {et~al.}(2015){Malacaria}, {Klochkov}, {Santangelo}, \&
  {Staubert}}]{Malacaria_etal15}
{Malacaria}, C., {Klochkov}, D., {Santangelo}, A., \& {Staubert}, R. 2015,
  \aap, 581, A121

\bibitem[{{Meszaros} \& {Nagel}(1985)}]{MeszarosNagel_85}
{Meszaros}, P. \& {Nagel}, W. 1985, \apj, 298, 147

\bibitem[{{Mihara} {et~al.}(1991){Mihara}, {Ohashi}, {Makishima}, {Nagase},
  {Kitamoto}, \& {Koyama}}]{Mihara_etal91b}
{Mihara}, T., {Ohashi}, T., {Makishima}, K., {et~al.} 1991, \pasj, 43, 501

\bibitem[{{Mukherjee} \& {Bhattacharya}(2012)}]{MukhBhatt_12}
{Mukherjee}, D. \& {Bhattacharya}, D. 2012, \mnras, 420, 720

\bibitem[{{Mukherjee} {et~al.}(2013){Mukherjee}, {Bhattacharya}, \&
  {Mignone}}]{Mukherjee_etal13b}
{Mukherjee}, D., {Bhattacharya}, D., \& {Mignone}, A. 2013, \mnras, 435, 718

\bibitem[{{Mukherjee} {et~al.}(2014){Mukherjee}, {Bhattacharya}, \&
  {Mignone}}]{Mukherjee_etal14}
{Mukherjee}, D., {Bhattacharya}, D., \& {Mignone}, A. 2014, in European
  Physical Journal Web of Conferences, Vol.~64, European Physical Journal Web
  of Conferences, 2004

\bibitem[{{Mushtukov} {et~al.}(2015){Mushtukov}, {Suleimanov}, {Tsygankov}, \&
  {Poutanen}}]{Mushtukov_etal15a}
{Mushtukov}, A.~A., {Suleimanov}, V.~F., {Tsygankov}, S.~S., \& {Poutanen}, J.
  2015, \mnras, 447, 1847

\bibitem[{{Orlandini} {et~al.}(1998){Orlandini}, {Dal Fiume}, {Frontera}, {Del
  Sordo}, {Piraino}, {Santangelo}, {Segreto}, {Oosterbroek}, \&
  {Parmar}}]{Orlandini_etal98}
{Orlandini}, M., {Dal Fiume}, D., {Frontera}, F., {et~al.} 1998, \apjl, 500,
  L163

\bibitem[{{Parmar} {et~al.}(1980){Parmar}, {Sanford}, \&
  {Fabian}}]{Parmar_etal80}
{Parmar}, A.~N., {Sanford}, P.~W., \& {Fabian}, A.~C. 1980, MNRAS, 192, 311

\bibitem[{{Petterson}(1977)}]{Petterson_77}
{Petterson}, J.~A. 1977, ApJ, 218, 783

\bibitem[{{Postnov} {et~al.}(2013){Postnov}, {Shakura}, {Staubert},
  {Kochetkova}, {Klochkov}, \& {Wilms}}]{Postnov_etal13}
{Postnov}, K., {Shakura}, N., {Staubert}, R., {et~al.} 2013, \mnras, 435, 1147

\bibitem[{{Postnov} {et~al.}(2015){Postnov}, {Gornostaev}, {Klochkov},
  {Laplace}, {Lukin}, \& {Shakura}}]{Postnov_etal15}
{Postnov}, K.~A., {Gornostaev}, M.~I., {Klochkov}, D., {et~al.} 2015, \mnras,
  452, 1601

\bibitem[{{Revnivtsev} \& {Mereghetti}(2016)}]{RevnivtsevMereghetti_16}
{Revnivtsev}, M. \& {Mereghetti}, S. 2016, {Magnetic Fields of Neutron Stars in
  X-Ray Binaries}, Vol.~54 (Springer 2016), 299

\bibitem[{{Reynolds} {et~al.}(1997){Reynolds}, {Quaintrell}, {Still}, {Roche},
  {Chakrabarty}, \& {Levine}}]{Reynolds_etal97}
{Reynolds}, A.~P., {Quaintrell}, H., {Still}, M.~D., {et~al.} 1997, \mnras,
  288, 43

\bibitem[{{Rodes-Roca} {et~al.}(2008){Rodes-Roca}, {Torrej{\'o}n}, \&
  {Bernab{\'e}u}}]{Rodes-Roca_etal08}
{Rodes-Roca}, J., {Torrej{\'o}n}, J.~M., \& {Bernab{\'e}u}, J.~G. 2008, {The
  fundamental cyclotron line in 4U 1538-52}, Vol.~3, 189--200

\bibitem[{{Rothschild} {et~al.}(2017){Rothschild}, {K{\"u}hnel}, {Pottschmidt},
  {Hemphill}, {Postnov}, {Gornostaev}, {Shakura}, {F{\"u}rst}, {Wilms},
  {Staubert}, \& {Klochkov}}]{Rothschild_etal17}
{Rothschild}, R.~E., {K{\"u}hnel}, M., {Pottschmidt}, K., {et~al.} 2017,
  \mnras, 466, 2752

\bibitem[{{Sazonov} {et~al.}(2020){Sazonov}, {Paizis}, {Bazzano}, {Chelovekov},
  {Khabibullin}, {Postnov}, {Mereminskiy}, {Fiocchi}, {B{\'e}langer}, {Bird},
  {Bozzo}, {Chenevez}, {Del Santo}, {Falanga}, {Farinelli}, {Ferrigno},
  {Grebenev}, {Krivonos}, {Kuulkers}, {Lund}, {Sanchez-Fernand ez}, {Tarana},
  {Ubertini}, \& {Wilms}}]{Sazonov_etal20}
{Sazonov}, S., {Paizis}, A., {Bazzano}, A., {et~al.} 2020, arXiv e-prints,
  arXiv:2006.05063

\bibitem[{{Schandl} \& {Meyer}(1994)}]{SchandlMeyer_94}
{Schandl}, S. \& {Meyer}, F. 1994, A\&A, 289, 149

\bibitem[{{Sch{\"o}nherr} {et~al.}(2007){Sch{\"o}nherr}, {Wilms}, {Kretschmar},
  {Kreykenbohm}, {Santangelo}, {Rothschild}, {Coburn}, \&
  {Staubert}}]{Schoenherr_etal07}
{Sch{\"o}nherr}, G., {Wilms}, J., {Kretschmar}, P., {et~al.} 2007, \aap, 472,
  353

\bibitem[{{Schwarm} {et~al.}(2017){Schwarm}, {Sch{\"o}nherr}, {Falkner},
  {Pottschmidt}, {Wolff}, {Becker}, {Sokolova-Lapa}, {Klochkov}, {Ferrigno},
  {F{\"u}rst}, {Hemphill}, {Marcu-Cheatham}, {Dauser}, \&
  {Wilms}}]{Schwarm_etal17a}
{Schwarm}, F.-W., {Sch{\"o}nherr}, G., {Falkner}, S., {et~al.} 2017, \aap, 597,
  A3

\bibitem[{Staubert(2003)}]{Staubert_03}
Staubert, R. 2003, in Multifrequency behaviour of high energy cosmic sources,
  ed. L.S.-G.F. Giovanelli, Vol. ChJAA, Vol. 3, S270

\bibitem[{{Staubert}(2014)}]{Staubert_14}
{Staubert}, R. 2014, in PoS(INTEGRAL2014)024

\bibitem[{{Staubert} {et~al.}(1983){Staubert}, {Bezler}, \&
  {Kendziorra}}]{Staubert_etal83}
{Staubert}, R., {Bezler}, M., \& {Kendziorra}, E. 1983, A\&A, 117, 215

\bibitem[{{Staubert} {et~al.}(2017){Staubert}, {Klochkov}, {F{\"u}rst},
  {Wilms}, {Rothschild}, \& {Harrison}}]{Staubert_etal17}
{Staubert}, R., {Klochkov}, D., {F{\"u}rst}, F., {et~al.} 2017, \aap, 606, L13

\bibitem[{{Staubert} {et~al.}(2013){Staubert}, {Klochkov}, {Vasco}, {Postnov},
  {Shakura}, {Wilms}, \& {Rothschild}}]{Staubert_etal13}
{Staubert}, R., {Klochkov}, D., {Vasco}, D., {et~al.} 2013, \aap, 550, A110

\bibitem[{{Staubert} {et~al.}(2016){Staubert}, {Klochkov}, {Vybornov}, {Wilms},
  \& {Harrison}}]{Staubert_etal16}
{Staubert}, R., {Klochkov}, D., {Vybornov}, V., {Wilms}, J., \& {Harrison},
  F.~A. 2016, \aap, 590, A91

\bibitem[{{Staubert} {et~al.}(2014){Staubert}, {Klochkov}, {Wilms}, {Postnov},
  {Shakura}, {Rothschild}, {F{\"u}rst}, \& {Harrison}}]{Staubert_etal14}
{Staubert}, R., {Klochkov}, D., {Wilms}, J., {et~al.} 2014, \aap, 572, A119

\bibitem[{{Staubert} {et~al.}(2007){Staubert}, {Shakura}, {Postnov}, {Wilms},
  {Rothschild}, {Coburn}, {Rodina}, \& {Klochkov}}]{Staubert_etal07}
{Staubert}, R., {Shakura}, N.~I., {Postnov}, K., {et~al.} 2007, A\&A, 465, L25

\bibitem[{{Staubert} {et~al.}(2019){Staubert}, {Tr{\"u}mper}, {Kendziorra},
  {Klochkov}, {Postnov}, {Kretschmar}, {Pottschmidt}, {Haberl}, {Rothschild},
  {Santangelo}, {Wilms}, {Kreykenbohm}, \& {F{\"u}rst}}]{Staubert_etal19}
{Staubert}, R., {Tr{\"u}mper}, J., {Kendziorra}, E., {et~al.} 2019, \aap, 622,
  A61

\bibitem[{{Tananbaum} {et~al.}(1972){Tananbaum}, {Gursky}, {Kellogg},
  {Levinson}, {Schreier}, \& {Giacconi}}]{Tananbaum_etal72}
{Tananbaum}, H., {Gursky}, H., {Kellogg}, E.~M., {et~al.} 1972, \apjl, 174,
  L143

\bibitem[{{Terada} {et~al.}(2007){Terada}, {Mihara}, {Nagase}, {Angelini},
  {Dotani}, {Enoto}, {Kitamoto}, {Kohmura}, {Kokubun}, {Kotani}, {Makishima},
  {Naik}, {Nakajima}, {Sugita}, {Sudoh}, {Suzuki}, {Takahashi}, {Yonetoku}, \&
  {Yoshida}}]{Terada_etal07}
{Terada}, Y., {Mihara}, T., {Nagase}, F., {et~al.} 2007, Adv. in Space
  Research, 40, 1485

\bibitem[{{Tr\"umper} {et~al.}(1986){Tr\"umper}, {Kahabka}, {Oegelman},
  {Pietsch}, \& {Voges}}]{Truemper_etal86}
{Tr\"umper}, J., {Kahabka}, P., {Oegelman}, H., {Pietsch}, W., \& {Voges}, W.
  1986, ApJ, 300, L63

\bibitem[{{Tr\"umper} {et~al.}(1978){Tr\"umper}, {Pietsch}, {Reppin}, {Voges},
  {Staubert}, \& {Kendziorra}}]{Truemper_etal78}
{Tr\"umper}, J., {Pietsch}, W., {Reppin}, C., {et~al.} 1978, ApJ, 219, L105

\bibitem[{Tr\"umper {et~al.}(1977)}]{Truemper_etal77}
Tr\"umper, J. {et~al.} 1977, Ann. N.Y. Acad. Sci., 302, 538

\bibitem[{Vasco(2012)}]{Vasco_12}
Vasco, D. 2012, PhD thesis, Univ. of T\"ubingen

\bibitem[{{Vasco} {et~al.}(2013){Vasco}, {Staubert}, {Klochkov}, {Santangelo},
  {Shakura}, \& {Postnov}}]{Vasco_etal13}
{Vasco}, D., {Staubert}, R., {Klochkov}, D., {et~al.} 2013, \aap, 550, A111

\bibitem[{{Voges} {et~al.}(1982){Voges}, {Pietsch}, {Reppin}, {Tr\"umper},
  {Kendziorra}, \& {Staubert}}]{Voges_etal82}
{Voges}, W., {Pietsch}, W., {Reppin}, C., {et~al.} 1982, ApJ, 263, 803

\bibitem[{{Vybornov} {et~al.}(2017){Vybornov}, {Klochkov}, {Gornostaev},
  {Postnov}, {Sokolova-Lapa}, {Staubert}, {Pottschmidt}, \&
  {Santangelo}}]{Vybornov_etal17}
{Vybornov}, V., {Klochkov}, D., {Gornostaev}, M., {et~al.} 2017, \aap, 601,
  A126

\bibitem[{Wilms(2012)}]{Wilms_12}
Wilms, J. 2012, in Proceed. 39th COSPAR Sci. Assembly, 14-22 July 2012, Mysore,
  India, Vol.~39, 2159

\bibitem[{{Wolff} {et~al.}(2016){Wolff}, {Becker}, {Gottlieb}, {F{\"u}rst},
  {Hemphill}, {Marcu-Cheatham}, {Pottschmidt}, {Schwarm}, {Wilms}, \&
  {Wood}}]{Wolff_etal16}
{Wolff}, M.~T., {Becker}, P.~A., {Gottlieb}, A.~M., {et~al.} 2016, \apj, 831,
  194

\bibitem[{{Xiao} {et~al.}(2019){Xiao}, {Ji}, {Staubert}, {Ge}, {Zhang},
  {Zhang}, {Santangelo}, {Ducci}, {Liao}, {Guo}, {Li}, {Zhang}, {Qu}, {Lu},
  {Li}, {Song}, {Xu}, {Bu}, {Cai}, {Cao}, {Chang}, {Chen}, {Chen}, {Chen},
  {Chen}, {Chen}, {Chen}, {Cui}, {Cui}, {Deng}, {Dong}, {Du}, {Fu}, {Gao},
  {Gao}, {Gao}, {Gu}, {Guan}, {Gungor}, {Guo}, {Han}, {Huang}, {Huo}, {Jia},
  {Jiang}, {Jiang}, {Jin}, {Kong}, {Li}, {Li}, {Li}, {Li}, {Li}, {Li}, {Li},
  {Li}, {Li}, {Liang}, {Liu}, {Liu}, {Liu}, {Liu}, {Liu}, {Lu}, {Lu}, {Luo},
  {Luo}, {Ma}, {Meng}, {Nang}, {Nie}, {Ou}, {Sai}, {Song}, {Sun}, {Tan}, {Tao},
  {Tuo}, {Wang}, {Wang}, {Wang}, {Wang}, {Wang}, {Wen}, {Wu}, {Wu}, {Wu},
  {Xiong}, {Yang}, {Yang}, {Yang}, {Yang}, {Yin}, {Yin}, {Zhang}, {Zhang},
  {Zhang}, {Zhang}, {Zhang}, {Zhang}, {Zhang}, {Zhang}, {Zhang}, {Zhang},
  {Zhang}, {Zhang}, {Zhang}, {Zhang}, {Zhang}, {Zhang}, {Zhao}, {Zhao},
  {Zheng}, {Zhou}, {Zhu}, \& {Zhu}}]{Xiao_etal19}
{Xiao}, G.~C., {Ji}, L., {Staubert}, R., {et~al.} 2019, Journal of High Energy
  Astrophysics, 23, 29

\end{thebibliography}
\end{document}